%% file: insync.tex
\documentclass[sigconf,screen,authorversion]{acmart}

\AtBeginDocument{%
  \providecommand\BibTeX{{%
    \normalfont B\kern-0.5em{\scshape i\kern-0.25em b}\kern-0.8em\TeX}}}

\input{meta/packages}
\input{meta/acronyms}
\input{meta/copyright}
\input{meta/definitions}

\input{meta/stats}

\usepackage{acmart-taps}

\begin{document}

\title[In Sync]{In Sync: Exploring Synchronization to Increase Trust Between Humans and Non-humanoid Robots}

\input{meta/authors}

\input{content/abstract}
\input{meta/meta}

\input{img/teaser}

\maketitle

\input{content/introduction}
\input{content/relatedwork}
\input{content/prototype}
\input{content/methodology}
\input{content/results}
\input{content/discussion}
\input{content/limitations}
\input{content/conclusion}

\input{meta/acks}

\bibliographystyle{ACM-Reference-Format}
\bibliography{insync_clean}

\end{document}

%% file: meta/packages.tex
\usepackage[nolist,nohyperlinks,smaller]{acronym}
\usepackage{subcaption}
\usepackage{siunitx}
\usepackage{booktabs}
\usepackage{multirow}
\usepackage{graphicx}
\usepackage{csquotes}

%% file: meta/acronyms.tex
\begin{acronym}[UMLX]
	\acro{HMD}{head-mounted display}
	\acro{GUI}{graphical user interface}
	\acro{HUD}{head-up display}
	\acro{TCT}{task-completion time}
	\acro{SVM}{support vector machine}
	\acro{EMM}{estimated marginal mean}
	\acro{AR}{Augmented Reality}
	\acro{VR}{Virtual Reality}
	\acro{RTLX}{Raw Nasa-TLX}
	\acro{WUI}{Walking User Interface}
	\acro{CSCW}{Computer-Supported Cooperative Work}
	\acro{HCI}{Human-Computer Interaction}
	\acro{ABI}{Around-Body Interaction}
	\acro{SLAM}{Simultaneous Location and Mapping}
	\acro{ROM}{range of motion}
	\acro{EMG}{electromyography}
	\acro{TPA}{Trust between People and Automation}
	\acro{SNHR}{Simple Non-Humanoid Robot}
\end{acronym}

%% file: meta/copyright.tex
\copyrightyear{2023} 
\acmYear{2023} 
\setcopyright{acmlicensed}\acmConference[CHI '23]{Proceedings of the 2023 CHI Conference on Human Factors in Computing Systems}{April 23--28, 2023}{Hamburg, Germany}
\acmBooktitle{Proceedings of the 2023 CHI Conference on Human Factors in Computing Systems (CHI '23), April 23--28, 2023, Hamburg, Germany}
\acmPrice{15.00}
\acmDOI{10.1145/3544548.3581193}
\acmISBN{978-1-4503-9421-5/23/04}

%% file: meta/definitions.tex
\newcommand{\factorMove}{\textsc{movement}}

\newcommand{\factorMoveLvlSimple}{\textsc{simple}}
\newcommand{\factorMoveLvlRandom}{\textsc{random}}
\newcommand{\factorMoveLvlSynchronized}{\textsc{synchronized}}

\newcommand{\dvMoney}{\textsc{money given to the prototype}}
\newcommand{\dvTPA}{\textsc{\ac{TPA}}}

\newcommand{\ivMovement}{\textsc{movement type}}

\newcommand{\trustGameMoney}{3\$}

\newcommand{\simNonHumRobot}{\textsc{\ac{SNHR}}}
\newcommand{\simNonHumRobots}{\textsc{\acp{SNHR}}}

%% file: meta/stats.tex
\newcommand{\subEtaG}[2]{%
	\ifthenelse{\equal{#1}{\string >.05}}
	{}
	{, $\eta_{G}^{2}=#2$}%
}

\newcommand{\subEta}[2]{%
	\ifthenelse{\equal{#1}{\string >.05}}
	{}
	{, $\eta^{2}=#2$}%
}

\newcommand{\efETAsquared}[1]{%
	\ifdim#1pt>0.139pt 
		large (\etasquared{} = #1)
	\else 
		\ifdim#1pt>0.059pt 
		medium (\etasquared{} = #1)
		\else 
		small (\etasquared{} = #1)
		\fi
	\fi
}

\newcommand{\kruskalwallis}[3]{$\chi^2(#1) = #2$, $p #3$}

\newcommand{\ztest}[2]{$z=#1, p#2$}

\newcommand{\val}[2]{$\mu = \num[round-mode=places,round-precision=1]{#1}$, $\sigma = \num[round-mode=places,round-precision=1]{#2}$}

\newcommand{\pearson}[2]{$r = #1$, $p#2$}

\def\etasquared{$\eta_{}^{2}$}

%% file: meta/authors.tex
\author{Wieslaw Bartkowski}
\orcid{0000-0002-2627-6529}
\affiliation{%
	\institution{University of Warsaw}
	\city{Warsaw}
	\country{Poland}
}
\email{wieslaw.bartkowski@uw.edu.pl}

\author{Andrzej Nowak}
\orcid{0000-0001-9965-0684}
\affiliation{%
	\institution{University of Warsaw}
	\city{Warsaw}
	\country{Poland}
}
\email{andrzejn232@gmail.com}

\author{Filip Ignacy Czajkowski}
\orcid{0000-0001-5618-6842}
\affiliation{%
	\institution{University of Warsaw}
	\city{Warsaw}
	\country{Poland}
}
\email{f.czajkowski@student.uw.edu.pl}

\author{Albrecht Schmidt}
\orcid{0000-0003-3890-1990}
\affiliation{%
	\institution{LMU Munich}
	\city{Munich}
	\country{Germany}
}
\email{albrecht.schmidt@um.ifi.lmu.de}

\author{Florian Müller}
\orcid{0000-0002-9621-6214}
\affiliation{%
	\institution{LMU Munich}
	\city{Munich}
	\country{Germany}
}
\email{florian.mueller@um.ifi.lmu.de}

\renewcommand{\shortauthors}{Bartkowski et al.}

%% file: content/abstract.tex
\begin{abstract}
	
When we go for a walk with friends, we can observe an interesting effect: From step lengths to arm movements - our movements unconsciously align; they synchronize. Prior research found that this synchronization is a crucial aspect of human relations that strengthens social cohesion and trust. Generalizing from these findings in synchronization theory, we propose a dynamical approach that can be applied in the design of non-humanoid robots to increase trust. We contribute the results of a controlled experiment with 51 participants exploring our concept in a between-subjects design. For this, we built a prototype of a simple non-humanoid robot that can bend to follow human movements and vary the movement synchronization patterns. We found that synchronized movements lead to significantly higher ratings in an established questionnaire on trust between people and automation but did not influence the willingness to spend money in a trust game.

\end{abstract}

%% file: meta/meta.tex
\ccsdesc[500]{Human-centered computing~Empirical studies in HCI}
\ccsdesc[300]{Human-centered computing~Laboratory experiments}
\ccsdesc[300]{Computer systems organization~Robotics}

\keywords{synchronization, trust, non-humanoid robot, dynamical approach, design strategy}

%% file: img/teaser.tex

\begin{teaserfigure}
	\includegraphics[width=\textwidth]{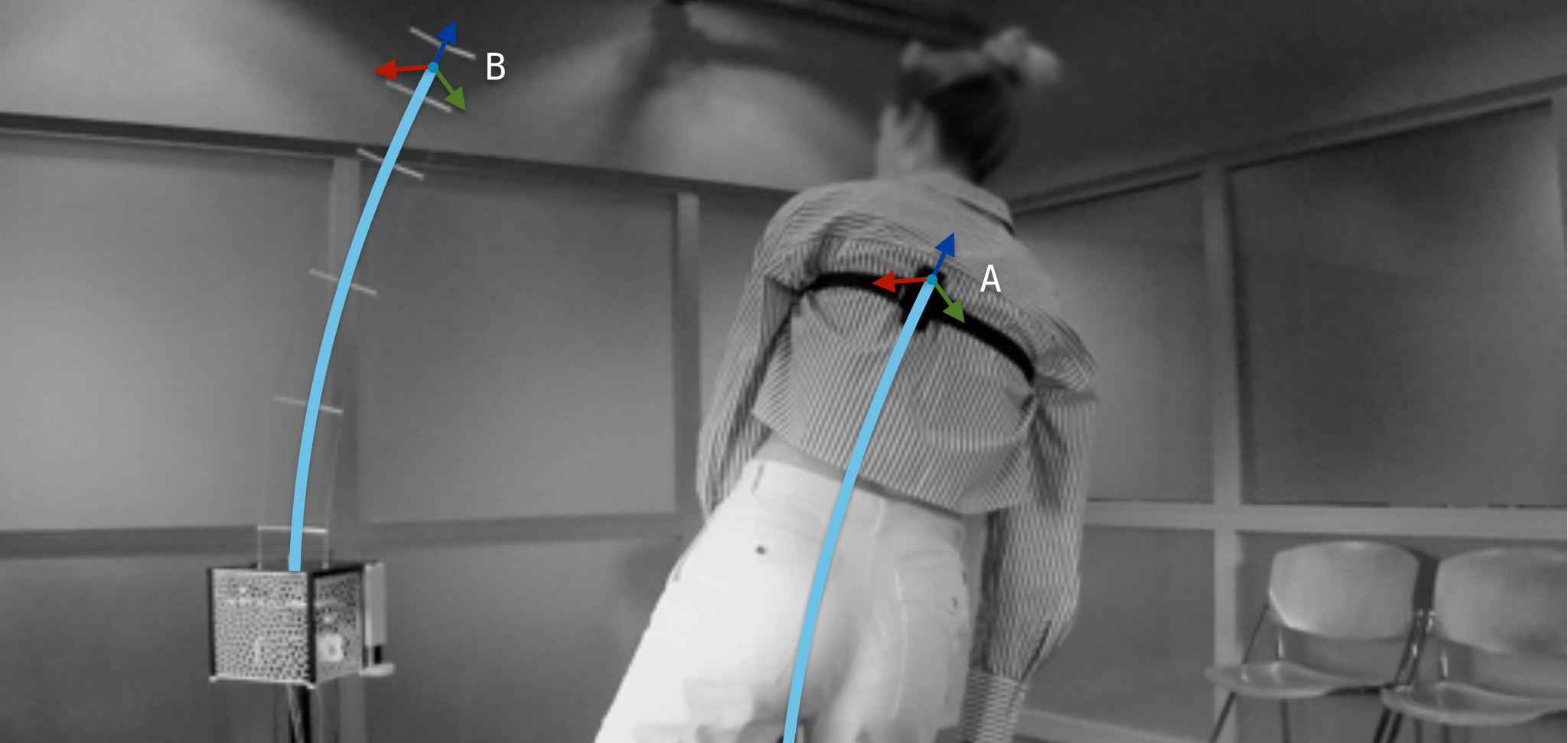}
	\caption{In this work, we investigate motion synchronization as a means of establishing trust between humans and simple non-humanoid robots. We built a prototype that can track and synchronize with the upper body movements of humans. The orientation of the continuum robot tip (B) is transformed according to changes in the orientation of the sensor attached to the participant's back (A).}
	\Description{Frame from video camera showing the free exploratory phase of the experiment. On the left prototype bends to the right, and on the right person interacting with the prototype bends to the right. On the picture market, two lines show how the orientation of the continuum robot tip is transformed according to changes in the orientation of the sensor attached to the participant's back.}
	\label{fig:teaser}
\end{teaserfigure}

%% file: content/introduction.tex
\section{Introduction}
\label{sec:introduction}

From entertainment~\cite{Pasquali2021, Aaltonen2017} or health care~\cite{Olaronke2017, Tuisku2019} to sex~\cite{Troiano2020, CoxGeorge2018, Richardson2016, Su2019}: with robots evolving from the assembly line to social robots, they are increasingly becoming part of our everyday lives. Thus, the question of how people perceive robots and what attitudes they have toward robots becomes crucial to define our relationships with them~\cite{naneva2020systematic}. Research has shown that trust is one of the key factors influencing the quality of interactions of humans with robots~\cite{safety8030048}, including a user's willingness to interact with a robot, take its advice into account~\cite{Langer2019}, and delegate tasks to robots~\cite{Hancock2011, Salem2015}. Lack of trust, hence, strains our relationship with social robots and may thus hinder or impede their future proliferation as ubiquitous everyday helpers.

In recent years, research has explored strategies by which robots can gain and maintain the trust of human users. This research concentrates mostly on robots’ features that influence how they interact with human users, pointing out such features as the capacity to reason~\cite{Kok2020}, the realistic facial expression of emotions~\cite{Salem2015}, or personalization~\cite{Langer2019}. \citet{kirkpatrick201710}  found that trust toward robots requires that the robot is perceived as having core human features such as agency and intention. Therefore, the more human a robot appears to be, the more trustworthy it can appear. If a robot displays human characteristics, it is likely to be perceived in interaction, in an anthropomorphic way, as a human~\cite{Langer2019, deVisser2017}. Building on this, research proposed a variety of human-like robot systems that can, for example, accurately express emotions through facial expressions or make eye-contact. The strategy of maximally anthropomorphizing robots, however, has its limitations. First, it raises unrealistic expectations that the robot will behave in a fully human way, which ultimately leads to frustration~\cite{Welge2016}. Second, it leaves aside the question of trust toward robots that do not have a humanoid appearance and advanced capacities for interaction with humans (e.g., reasoning). We refer to such kind of robot as an \simNonHumRobot{}. Simple as not having advanced capacities for interaction, and non-humanoid as not having a humanoid appearance.

To overcome these limitations, in this work, we go beyond state-of-the-art and add to the body of research in human-robot interaction by exploring a novel strategy to establish trust between humans and \simNonHumRobots{}. Drawing on recent findings in social psychology and neuropsychology, we propose to leverage the effect of synchronization: When we go for a walk with friends, we can observe that our movements - from stride length to arm movements - unconsciously align; they synchronize. This synchronization is a critical feature of positive social relations~\cite{Semin2008, Nowak2017, Nowak2020} and trust~\cite{Cheng2022}. 


Generalizing the findings in social psychology, we hypothesize that the synchronization of an \simNonHumRobot{} with the human user increases the trust toward the \simNonHumRobot{}. Because synchronization is a basic dynamical feature of interaction, it does not assume a humanoid shape of a robot and advanced capacities for interaction. As a first step to evaluate the feasibility of establishing such a physical synchronization between humans and robots, we explore our idea using an intentionally abstract object looking more like an art installation than a robot to exclude possible confounding factors. For conducting the study, we built a prototype implementation of such an object, which tracks the upper body movements of people in its vicinity and can synchronize its movements with their movements (see fig \ref{fig:teaser}).

The contribution of this paper is two-fold. First, we illustrate the design process and technical implementation of a robot prototype that allows physical synchronization with human users. Second, we contribute the results of a controlled experiment with 51 participants assessing the influence of synchronized movement compared to simple or random movement patterns of a robot on the perceived trust of users and propose a set of guidelines for the future use of such interfaces.

%% file: content/relatedwork.tex
\section{Related Work}
\label{sec:relatedwork}

Our work was heavily influenced by a large body of prior works in the areas of synchronization and trust in human-robot interaction and synchronization and trust in humans.

\subsection{Trust in Human-Robot Interaction}


Trust is the primary factor shaping social interactions~\cite{doi:10.1080/21515581.2012.708496}, from intimate relations~\cite{Rosenthal1981}, through work relations~\cite{Brown2015} and consumer behavior~\cite{Ding2013} to conflict~\cite{Malhotra2011}. Most definitions of trust rely on the notion that trust describes a relation between a trustor (subject) and a trustee (object) that are interdependent in the sense that the action of one has some consequences for the other in a situation containing risks for the trustor \cite{pytlikzillig2016consensus}. Further, trust can be understood as a form of reliance that is based on the judgment that the partner has the relevant competence, motivation, and opportunity~\cite{de2021defining}.

Following the pivotal role of trust in interpersonal relationships, trust has also emerged as an increasingly relevant area of research in human-robot interaction~\cite{Hancock2011,gompei2018factors,hancock2021evolving,baker2018toward,plaks2022identifying}. In this field, many studies explored the antecedents of trust toward robots. They found that trust toward robots depends on human personality factors, features of the situation, history of interactions with robots, and characteristics of the robots~\cite{Hancock2011,Kok2020,naneva2020systematic}. Further, the research found that previous interactions with robots result in higher trust \cite{sanders2017trust, correia2016just}. However, these effects of prior experience with robots depend on how the robot behaves in the interaction. In a meta-analysis of trust in HRI, \citet{Hancock2011} found that robot performance-based factors (e.g., reliability, false-alarm rate, failure rate) had the greatest influence on developing trust in the robot. As another perspective on trust toward robots, \citet{gompei2018factors} explored cognitive and affective trust as two dimensions of trust toward robots. Cognitive trust is semi-rational and strongly influenced by the perception of the reliability of the robot, its intention, and its goals. On the other hand, affective trust is emotional in nature and less affected by the robot’s mistakes than cognitive trust but by the robot’s appearance (human-like appearance leads to more trust). As an example of such affective trust, it recently demonstrated that robots with music-driven emotional prosody and gestures were perceived as more trustworthy than robots without such features \citet{savery2019establishing}.

Recently, we have seen a shift in focus of the research on trust in robotics, from studying the causes and effects of trust in robots to the strategies for robots to actively gain and maintain the trust of humans. Most of this work concentrates on two domains, making robots more humanoid with increased social characteristics such as emotional facial expression ~\cite{calvo2020effects} or increasing their reasoning capacity ~\cite{naneva2020systematic,Salem2015,christoforakos2021can}. These directions, however, have important limitations as they do not transfer to non-humanoid and simple robots, which do not resemble humans and are not capable of verbal communication and high reasoning capacity. Therefore, this work focuses on strategies to establish trust toward \aclp{SNHR}.



\subsection{Synchronization in Humans and Human-Robot Interaction}


In physics, synchronization is the alignment of the rhythms of two or more oscillators due to mostly weak interactions~\cite{pikovsky2001synchronization} and, therefore, the coordination in time among the states or dynamics of the elements comprising the system \cite{schmidt2008dynamics}. Beyond physics, the notion of synchronization has been adapted for other systems, such as periodic dynamics in cardiac~\cite{Rosenblum1998} or nervous systems~\cite{Lestienne2001}. \citet{pikovsky2001synchronization} provide a description of specific types of synchronization. The most prominent type of synchronization is mimicry \cite{chartrand2009human}, also known as the chameleon effect \cite{chartrand1999chameleon}. In mimicry, one interaction partner mimics with some time delay and possibly some variation, movements, postures, rate of speech, or the tone of voice of the other interaction partner.

Similarly to the usage in other domains, synchronization plays a key role in social interactions~\cite{yun2012interpersonal,Nowak2017,Nowak2020}. In such social interactions, synchronization of behavior binds individuals into higher-level functional units, such as dyads or social groups~\cite{Nowak2017,Nowak2020}, that can perform a common action aimed at the achievement of a common goal, for example, moving a piece of furniture, servicing a car in a race, prepare a meal in a picnic or singing a song together. For the interaction between two individuals to proceed smoothly, the overt behavior and internal states (e.g., activation level, emotions, goals) of the individuals must achieve synchronization \cite{newtson1994perception,fusaroli2014dialog}. This synchronization needs to occur at various levels, including motor behavior, but it also involves cognitive and emotional dynamics \cite{nowak2000modeling}. Synchronization does not need to be the same in all modalities. Two individuals involved in a conversation, for example, in an antiphase manner, synchronize their speech while they synchronize their facial expressions and body movements in-phase manner \cite{stel2010mimicry}. In this context, antiphase means that when the former speaks, the latter is silent, and vice versa. In-phase means that the former's body movements and facial expressions are in unison with the latter's body movements and facial expressions. The synchronization on the emotional level is related to the synchronization on the behavioral level because facial expressions tend to induce the corresponding emotional state in each partner of the conversation \cite{laird1974self,strack1988inhibiting}. Interpersonal synchronization protects against the antisocial consequences of frustration \cite{dybowski2022interpersonal}. Further, synchronization has important consequences on social relations  \cite{baron1994local, newtson1994perception, nowak1998computational, marsh2009social, miles2009rhythm}. Synchronization other leads to the formation of social ties and promotes a feeling of connectedness and liking \cite{chartrand1999chameleon, lakin2003using, dijksterhuis2005we, hove2009s}, while the failure to achieve synchronization evokes feelings of separateness \cite{nowak2007dynamical}. Highly related, \citet{launay2013synchronization} found that synchronization increases trust toward others \cite{launay2013synchronization}. Further, recent works indicated that synchronization of group decisions resulted in higher trust \cite{daudi2016effects}, and synchronization on the brain level in a trust game is correlated with higher investment and, thus, higher trust \cite{Cheng2022}.

Further, there is a large body of work in the area of HCI on human interaction mediated by technology, exploring mediated trust and mediated synchronization. For example, \citet{bos2002effects} found trust formation was slower, and the trust achieved was more fragile in mediated conditions. \citet{riegelsberger2003researcher} propose a methodological foundation to assess mediated trust. \citet{slovak2011exploring} suggest that even subtle differences in video-conference design may significantly impact mediated trust. \citet{brave1998tangible} explore the physical synchronization of states of distant, identical objects to create the illusion of shared physical objects across distance. \citet{rinott2022designing} indicate the potential of using interpersonal motor synchronization in HCI thanks to its pro-social consequences. \citet{slovak2014exploring} connect changes in skin conductance synchrony to changes in emotional engagement. \citet{scissors2008linguistic} demonstrated that forms of linguistic mimicry are associated with establishing trust between strangers in a text-chat environment. In this following, we specifically focus on HRI and the relationship between synchronization and trust.


In recent years, research started to explore synchronization in human-robot interaction. For example, \citet{hofree2014bridging} found that humans will mimic the facial expressions of a robot, even if they are fully aware that their interaction partner is non-human. \citet{shen2015can} found that the motor coordination mechanism improved humans' overall perception of the humanoid robot. Other work has explored robotic drumming, rhythmic HRI, and human-robot musical synchronization \cite{crick2006synchronization,hoffman2011interactive,weinberg2009leader}. \citet{mortl2014rhythm} developed the concept of goal-directed synchronization behavior for robotic agents in repetitive joint action tasks and implemented it in an anthropomorphic robot.
Further, \citet{hashimoto2009effects} explored emotional synchronization and found that human feelings became comfortable when the robot made the synchronized facial expression with human emotions. Further works, again involving humanoid robots, found that the presence of a physical, embodied robot enabled more interaction, better drumming, and turn-taking, as well as enjoyment, especially when the robot used gestures~\cite{kose2009effects}. However, to the best of our knowledge, there exists no prior work exploring synchronization as a means to increase trust between humans and \simNonHumRobots{}. 

%% file: content/prototype.tex
\input{img/prototype_img}

\section{Design Considerations and Prototype}
\label{sec:prototype}

The question which motivates our research is how to design \simNonHumRobots{} so humans will trust them. The answer to this question is relatively straightforward with respect to the cognitive dimension – the robot should look reliable and should not make errors. However, this question is much more interesting with respect to the affective dimension of trust, which arguably is even more important because affective trust increases the willingness to cooperate and forgive errors. Research on trust in humans suggests that the ability to synchronize with the human partner may be an important factor in perceiving the robot as trustworthy.

Based on the above motivation and a review of related work, in this paper, we explore synchronization as a means of establishing trust between humans and \simNonHumRobots{}. While the literature highlights various types of body movements and body signals that could be used to establish such synchronization, in our work, we explore synchronization with upper body movements as an example. This way, we wanted to avoid complex movements on the participant's side, such as gesticulating or nodding. We wanted to obtain the simplest possible patterns leading to synchronization at the level of whole-body movement. For this, we designed a prototype that can mimic the participant’s upper body movements by bending the prototype's upper part. Besides synchronized movements, it can also generate random or simple movement patterns. In the following section, we present the design goals and the implementation of the prototype.

\subsection{Design Considerations}

To exclude possible confounding factors, we opted for an abstract form. That is, the participant would not associate it with anything known. In particular, we wanted to avoid anthropomorphizing the form of the robot. We also tried to prevent associations with all animate forms, which could be associated with expectations of preserving the prototype or carrying specific attitudes or emotions. We also avoided association with known machines, particularly with all kinds of robots, including industrial robots. For this reason, we chose to construct a 3-tendon single-segment continuum robot \cite{Robinson1999ContinuumR} that has no rigid joints associated with a stereotypical robot. Additionally, in the participant’s instructions, we used the word installation instead of a machine or robot to direct associations toward an abstract art object rather than a practical device.



\subsection{The prototype}
\label{sec:prototype:construction}

We chose a simple form, a flexible vertical element with a length of 1 meter, placed on a 90 cm height stand. The movement is obtained by bending the flexible element into an arc with a variable radius and direction of deflection. In pilot studies, we noticed that this form of object and this type of movement made study participants tend to imitate the prototype's movement by flexing the upper body and flexing the spine. With increasing intensity of body movement, participants started to move their hips and shoulders. 

We prepared a GitHub repository\footnote{GitHub repository: https://github.com/wbartkowski/In-Sync-Robot-Prototype} containing the prototype's technical documentation, mechanical parts, 3D models for printing, electronic schematics, software source code, and a complete list of required components from external vendors. We hope it helps the HCI community, e.g., replicate the experiment or build the robot for other purposes. 


\subsubsection{Construction, mechanics, materials, actuation, and safety}

We opted to use a cable/tendon actuation, the most common method of driving continuum robots. As depicted in fig. \ref{fig:insync:prototype:model}, our cable-driven continuum robot has multiple spacer structures connected in series by a backbone located at the center axis. The three cables are spaced 120 degrees around the center backbone. Cables pass through a series of spacer structures that keep the cables in the correct position relative to the backbone. The end of the cables is fixed to the end structure on the top and to the actuators on the other end. As actuators pull the cable, the cable length inside the structure of the continuum robot is decreased, thus forcing the robot's structure to bend toward the side of the pulled cable. Through the coordinated displacement of each driving cable, the continuum robot can bend toward any specified direction \(\theta\) with a defined bending angle \(\phi\). As actuators    , we used two-phase stepper motors 17HS4401 controlled by TRINAMIC's TMC2209, an ultra-silent motor driver. Using this setup, we managed to achieve a smooth and noiseless movement, thus eliminating additional factors that may affect the perception of the study participant.

For safety reasons, we have limited the bending angle to 20 degrees to prevent the robot's tip from hitting the participant if they got too close. Because some delay is needed in mimicry, we also have limited acceleration to \(62.8\frac{cm}{s^2}\) and a maximum speed of cable pulling to \(25.12\frac{cm}{s}\).

We built two iterations of the prototype based on different materials. The first prototype uses a backbone made of a densely wound spring steel spring with a diameter of 12mm. Pilot studies indicated that participants perceived this as heavy and dangerous. Additionally, some participants were concerned about being accidentally hit by the robot. A sense of insecurity or an unfriendly appearance eliminated the possibility of establishing trust. Therefore, in the second iteration, we used a 3mm diameter fiberglass backbone also used to construct kites. The participants perceived this structure as light, airy, and non-threatening.


Other crucial parts of the prototype's structure were designed in CAD software Fusion 360, and 3D printed using PET-G filament, having excellent mechanical properties.

\subsubsection{Electronic, connectivity, and software}

We use Espressif Systems ESP32 chip with Xtensa® 32-bit LX6 microprocessors to compute the robot movement formula and control the actuator drivers. We chose ESP32 because it supports the ESP-NOW protocol developed by Espressif, which enables multiple devices to communicate with one another using ESP32’s Wi-Fi hardware without needing a Wi-Fi router, reducing the complexity of the hardware setup. Moreover, ESP-NOW is connectionless with no handshake required (as required for, e.g., Bluetooth pairing), so this solution was very convenient for our research setting, where three devices (robot prototype, orientation sensor device, and data recorder) have to communicate constantly with low latency and the possibility of instant reconnection in case of accidental power loss and during research hardware setup. 
 
Additionally, in the prototype, we use a separate microcontroller based on ATmega328P to control experiment conditions and coin acceptor. ATmega328P is communicating with ESP32 over the USART interface.  

Software for the prototype and other devices is written in C language and developed using Visual Studio Code environment with PlatformIO extension for easy management of different electronics development platforms.

\input{img/prototype_schematic_math}

\subsection{Movement Tracking and Mapping to Movements of the Prototype}

\subsubsection{Custom-built orientation sensor}

We tracked the participants' upper body movements using a custom-built orientation sensor device based on the Bosch BNO055 9-DOF IMU sensor with integrated sensor data fusion algorithms and calibration algorithms. We used the ESP-NOW protocol for wireless communication with the prototype based on the ESP32 chip (see the previous section). We integrated into the device a lithium-ion battery with a charger. Using a fusion algorithm, we extracted the orientation data, such as a pitch, roll, and heading, from the sensor. We transformed the orientation data through the formula (see fig.~\ref{fig:insync:prototype:math}) into the bending as described in the next subsection, \ref{sec:mapping}, where the heading corresponds to \(\theta\), and pitch and roll are used to compute \(\phi\). 

\subsubsection{Movement mapping}
\label{sec:mapping}

The orientation of the sensor placed on the participant's back is transformed into the orientation of the reference system at the tip of the prototype backbone (see fig. \ref{fig:teaser}) by bending the backbone caused by stretching the cables running along the inextensible backbone (see fig. \ref{fig:insync:prototype:model}). The degree of stretching of the cables \(l_1, l_2, l_3\) is calculated according to the formula~\cite{nazari2019forward} derived from the bending parameters of the curve defined by the bend direction angle \(\theta\) and bend angle \(\phi\) (see fig. \ref{fig:insync:prototype:mathsche}). 

%% file: img/prototype_img.tex
\begin{figure*}[ht!]
 \centering
\includegraphics[width=\textwidth]{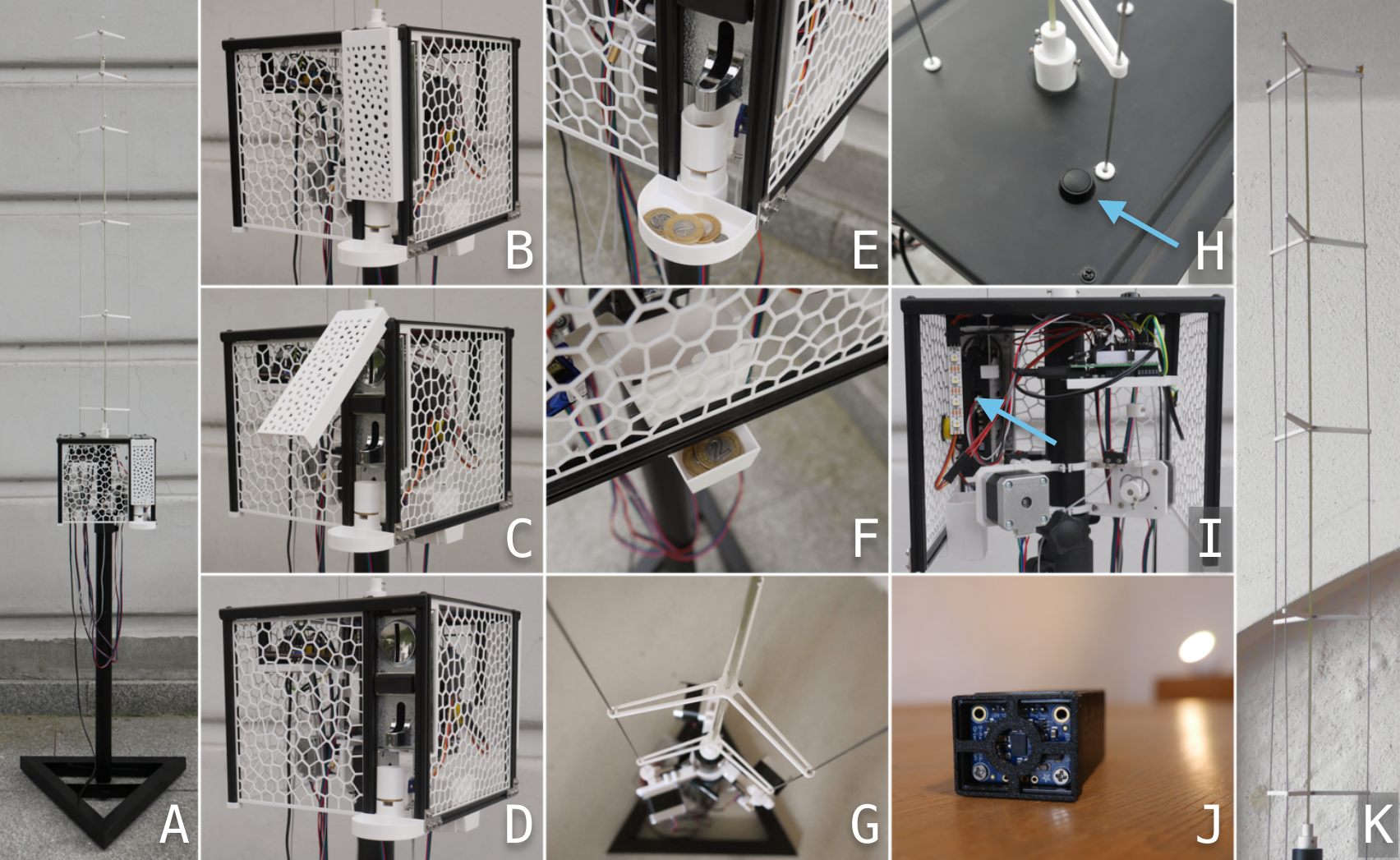}
\caption{Prototype design details. (A) The prototype’s general view. We see starting from the top: a single-segment continuum robot; a body of the prototype with electronics, actuators, and a coin acceptor; a plinth in the form of a pipe with a triangular foot. (B) The body with a covered coin acceptor. (C) Coin acceptor cover opening. (D) A body with a visible coin acceptor. (E) A coin returner with a coin cup refilling before each experiment. Below the tray where the returned coins fall. (F) A hidden drawer in which the coins thrown by the participant are collected. Coins are taken out after each experiment. (G) Zooming in on the elements of a single-segment continuum robot construction. Visible: fiberglass backbone, tendons made of braided flexible steel lines, separators made of PET-G, keeping the tendons at a proper distance from the backbone. (H) The experiment stage change button, operated by the investigator, is unnoticeable to the participant. (I) The arrow points to the indicator of condition number and stage of the experiment. The indicator facilitates the operation by the investigator. In addition, the photo shows the actuators and part of the control electronics. (J) IMU sensor to place on the participant's back during the exploratory phase of the experiment. (K) Zooming in on the single-segment continuum robot.}
\Description{Collage of 11 small pictures showing prototype design details. (A) The prototype’s general view. We see starting from the top: a single-segment continuum robot; a body of the prototype with electronics, actuators, and a coin acceptor; a plinth in the form of a pipe with a triangular foot. (B) The body with a covered coin acceptor. (C) Coin acceptor cover opening. (D) A body with a visible coin acceptor. (E) A coin returner with a coin cup was refilled before each experiment. Below the tray where the returned coins fall. (F) A hidden drawer in which the coins thrown by the participant are collected. Coins are taken out after each experiment. (G) Zooming in on the elements of a single-segment continuum robot construction. Visible: fiberglass backbone, tendons made of braided flexible steel lines, separators made of PET-G, keeping the tendons at a proper distance from the backbone. (H) The experiment stage change button operated by the investigator is unnoticeable to the participant. (I) The arrow points to the indicator of condition number and stage of the experiment. The indicator facilitates the operation by the investigator. In addition, the photo shows the actuators and part of the control electronics. (J) IMU sensor to place on the participant’s back during the exploratory phase of the experiment. (K) Zooming in on the single-segment continuum robot.}
\label{fig:insync:prototype}
\end{figure*}

%% file: img/prototype_schematic_math.tex
\begin{figure*}[t!]
 \centering
 \begin{subfigure}[b]{0.22\textwidth}
    \centering
    \includegraphics[width=\textwidth]{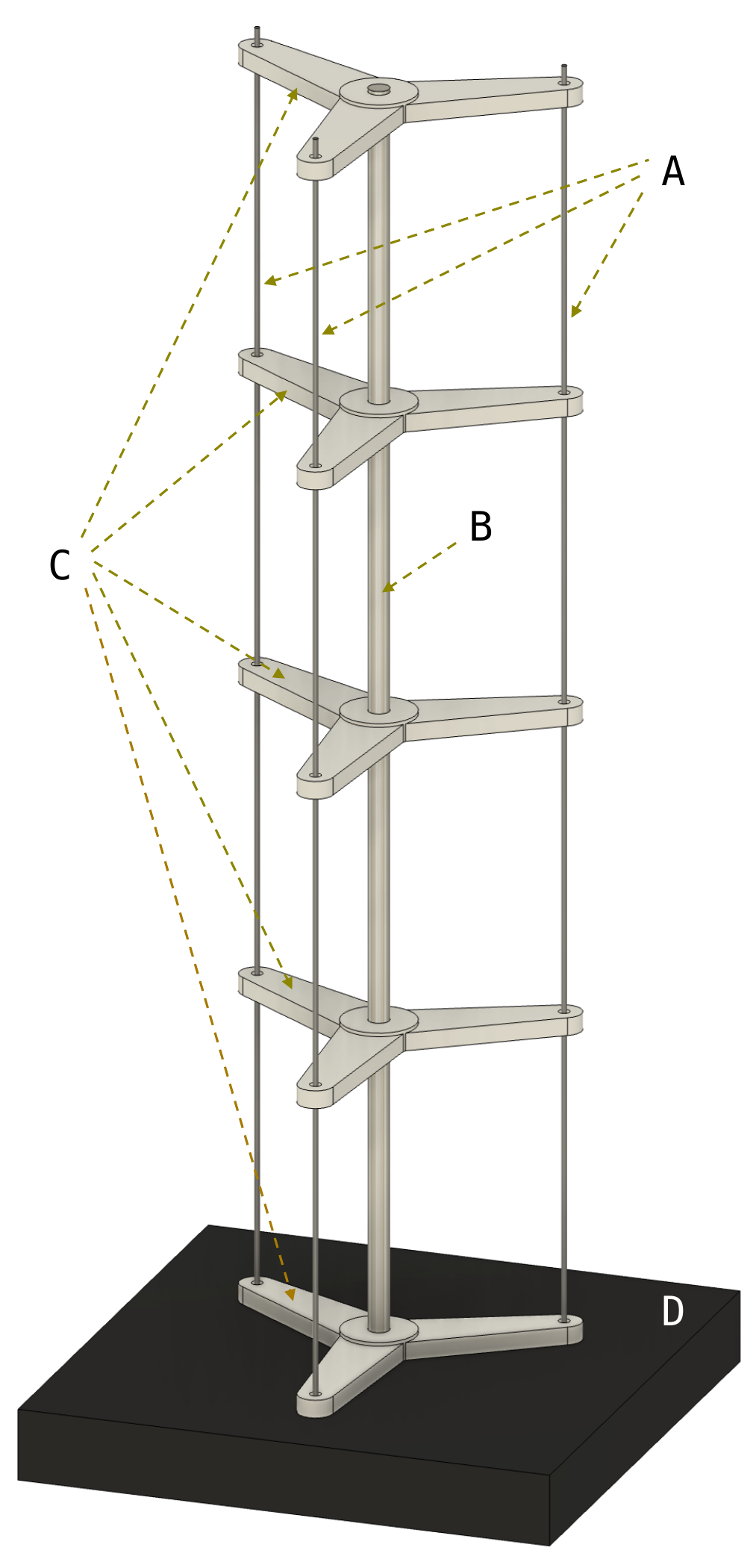}
    \caption{}
    \label{fig:insync:prototype:model}
\end{subfigure}
\hfill
\begin{subfigure}[b]{0.3\textwidth}
    \centering
    \includegraphics[width=\textwidth]{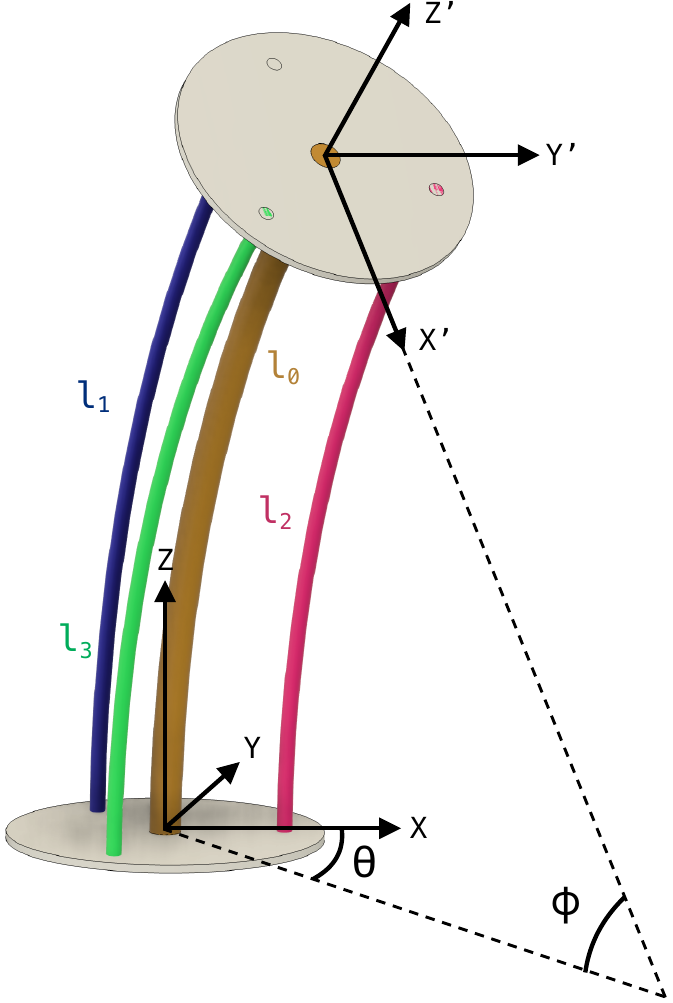}
    \caption{}
    \label{fig:insync:prototype:sche}
\end{subfigure}
\hfill
\begin{subfigure}[b]{0.3\textwidth}
    \centering
    \begin {align*}
        l_1 = (-r\cos \theta)\phi
        \\
        \\
        l_2 = \left(\frac{1}{2}r\cos \theta-\frac{\sqrt{3}}{2}r\sin \theta\right)\phi
        \\
        \\
        l_3 = \left(\frac{1}{2}r\cos \theta-\frac{\sqrt{3}}{2}r\sin \theta\right)\phi
    \end {align*}
    \caption{}
    \label{fig:insync:prototype:math}
\end{subfigure}
\caption{(a) The model of a one-section cable-driven continuum robot used in the prototype: A - three cables/tendons; B - inextensible backbone; C - five spacer structures; D - base. (b) The layout of a one-section cable-driven continuum robot used in kinematic computing. \(l_0\) length of inextensible backbone; \(l_1, l_2, l_3\) length of cables driven by actuators; \(\phi\) angle of backbone bending; \(\theta\) angle of backbone bending direction; \(x, y, z\) backbone base coordinate system; \(x', y', z'\) backbone tip  coordinate system. (c) The formula provides the length of cables. Based on that length, actuators shorten or lengthen cables accordingly.}
\label{fig:insync:prototype:mathsche}
\Description{A three-part figure: (a) The model of a one-section cable-driven continuum robot used in the prototype with marked parts: A - three cables/tendons; B - backbone; C - five spacer structures; D - base. (b) The layout of a one-section cable-driven continuum robot used in kinematic computing. With marked elements: l0 length of inextensible backbone; l1, l2, l3 length of cables driven by actuators; phi angle of backbone bending; theta angle of backbone bending direction; x, y, z backbone base coordinate system; xprime, yprime, zprime backbone tip coordinate system. (c) The formula provides the length of cables. Based on that length actuators shorten or lengthen cables accordingly.}
\end{figure*}

%% file: content/methodology.tex
\section{Methodology}
\label{sec:methodology}

Based on the proposed concept and prototype implementation, we conducted a controlled experiment to investigate the influence of motion synchronization on participants' trust formation toward an \simNonHumRobot{}. More specifically, we investigated the following research question:


\begin{description}
	\item[RQ] Does the synchronized movement of the robot change the feeling of trust toward an \simNonHumRobot{}?
\end{description}




In the following section, we report on the methodology of our experiment.

\subsection{Design and Task}
\label{sec:methodology:design}

To answer the research questions, we conducted a controlled experiment in which participants interacted with an \simNonHumRobot{}, as described in section \ref{sec:prototype}. We varied the movement pattern of the robot in a between-subjects design with three levels (two unsynchronized and one synchronized movement pattern). We explained to the participants that we were working on a prototype for human-machine interaction. To avoid biasing the participants, we did not give any further information about the purpose and goal of the prototype. We chose not to give our participants any specific tasks to perform or goals to achieve. Instead, we asked them to interact with the robot in a natural and spontaneous way without any constraints or instructions. For this, participants were allowed to move around the room at will for 3 minutes (the pilot studies showed that interest in the task lasted for about 3 minutes). We opted for this approach to assess the effects of synchronization on trust in a more natural and realistic setting without the potential confounds that would arise from using a specific task or context. More specifically, a task-oriented scenario would allow measuring the confidence in the machine (cognitive dimension, e.g., reliability, efficiency) but would not allow measuring the relationship with the machine (affective dimension, e.g., feelings, emotions). Moreover, a task-oriented scenario would make it difficult to manipulate the synchronization so that it does not affect the quality of the execution of the task. 

After this, we asked participants to fill out a questionnaire and handed them their compensation of about \trustGameMoney{} (in local currency) as five coins. As the last step, we gave our participants the option to optionally play a version of the trust game \cite{BERG1995122} with the prototype: Participants could deposit any portion of the coins they received (from 0 to 5 coins) into the prototype (if someone refused to participate in the game, we counted it as depositing 0 coins). We informed the participants that the prototype would be credited with the tripled amount of their deposit and subsequently, at will, would pay a share (i.e., between 0\% and 300\% of the inserted money) of this sum back.

Following the results of our literature review and informal pretests, we expected that the type of movement performed by the prototype during the interaction would affect the participants' sense of trust in interacting with the \simNonHumRobot{}. Therefore, we varied the \factorMove{} as an independent variable with three levels, namely:

\begin{description}
	\item[\factorMoveLvlSynchronized{}] as a prototype with movements synchronized to the participant's movements. The participant's movements are mimicked to achieve a specific form of synchronization - delayed in time and possibly spatially transformed. So participant movements are transformed into movements of the prototype using a formula for the robot's kinematics (see fig. \ref{fig:insync:prototype:math}) with added movement delay by limiting maximum speed and acceleration as described in section \ref{sec:prototype:construction} and adding a rotational transformation (by 10 degrees) of the bending direction \(\theta\).
	\item[\factorMoveLvlRandom{}] as a prototype with random movements using Brownian motion \cite{hida1980brownian} to program it. We chose the parameters of the Brownian motion in a way that the amplitude and frequency of the motions were comparable to the motions emitted by the prototype in synchronization mode.
	\item[\factorMoveLvlSimple{}] as a simple, recurring pattern of movement. We program it as sinusoidal, waving sideways in one plane. 
\end{description}

We varied the independent variable in a between-subjects design by assigning each participant to one of the three conditions. For each condition, we measured the following dependent variables:

\begin{description}
	\item[\dvTPA{}] To further gain insight into the trust relationship between the prototype and the participant, we employed the widely used \ac{TPA} checklist as proposed by \citet{Jian2000}. The \ac{TPA} consists of twelve items with 7-point Likert scales each. It measures trust and distrust as polar opposites along a single dimension. Therefore, the output may be a single all-encompassing trust value or separate values for the trust and distrust dimensions \citet{kohn2021measurement}.
	\item[\dvMoney{}] Similar to previous work in assessing trust in robots \cite{mota2016playing,ALARCON2023103858,oksanen2020trust,zorner2021immersive}, we used the amount of money staked in the trust game \cite{BERG1995122} as a measure of the trust participants placed in the prototype.
\end{description}

Based on the study design, we obtained ethics approval from our institution before the experiment, which had no objections.





\subsection{Study Setup and Apparatus}

For the experiment, we used the prototype as described in section \ref{sec:prototype}. We modified the prototype to support the trust game as described above. For this, we built a typical coin acceptor into the prototype, which, as expected, turned out to be a good affordance for the action of inserting coins into it. The coin acceptor also counted the number of inserted coins. The operation of the coin acceptor was coordinated with the operation of the coin returner made by us, which, 10 seconds after the participant had thrown in the last coin, started the process of returning an adequate number of coins by the prototype (see fig. \ref{fig:insync:prototype}D-E).

Further, we used a data recorder placed on the technical table in a far-away corner of the room (see fig. \ref{fig:insync:setup:plan}). We developed a data recorder device to ensure the quality of data collected on paper by the investigator. We recorded the following data: movement tracking data, number of inserted coins, research condition number, timestamp, and time of every experiment stage. For the construction, we used the same chip ESP32 (see section \ref{sec:prototype}) with added microSD card driver. For communication with the prototype, we used the ESP-NOW communication protocol (see section \ref{sec:prototype}).

Besides this, we also used Sony digital camera model no. DSC-HX5V for video recording captures general situation and participant movements for further analysis (see fig. \ref{fig:insync:methodology:studysetup}). In the same room, we also prepared a table for filing a consent form before the study and a questionnaire and compensation confirmation form after the study. We decided to have this table in the same room because after giving compensation, the moderator proposes a trust game with the prototype, so we want to have the prototype in close range all the time (see fig. \ref{fig:insync:setup:plan}).

We decided to put the prototype in the corner of the room, with the front rotated 45 degrees to the wall. This arrangement emphasizes the position of the front of the prototype while allowing the participant to walk around the prototype freely. We left space of 120 cm between the sides of the prototype and the side walls to make it possible. The initial position of the participant for each study is the same, i.e., 130 cm from the front of the prototype. See fig. \ref{fig:insync:setup:plan}, where the participant’s starting position is marked with a circle with a dashed brown line.

\input{img/study_setup}


\subsection{Procedure}


After welcoming the participants, we introduced them to the general topic of the experiment and led them to the room with the prototype and other stuff (see fig. \ref{fig:insync:methodology:studysetup}). Then, we asked them to fill out a consent form. After that, we read the following instructions: \emph{\enquote{The aim of the research is to help create machines controlled by artificial intelligence and other physical systems cooperating with humans. Here you can see installations controlled by artificial intelligence. Your task is to check how the installation reacts to your movements. In a moment, I will leave you with the installation so that you can freely explore its possibilities. You have 3 minutes to do so. Please do not touch the installation. After 3 minutes, I will come back and ask you to fill in the questionnaire in which you will evaluate the installation}}. We further instructed the participants that we would measure their physiological body signals and informed them that we would need to attach a recording device to their bodies for this purpose. In fact, we attached the orientation sensor device (see fig. \ref{fig:insync:prototype}J) as described in the previous section. We did not give the participants further instructions about their actions and let them freely explore the prototype in their chosen way. The investigator asked the participant to stand approximately 1.3 meters in front of the prototype (see fig. \ref{fig:insync:setup:plan}) and started video recording. Then investigator started the exploratory experiment phase and left the room, leaving the participant alone with the prototype.

After participants had 3 minutes to explore the prototype and its responses, the investigator returned to the room and stopped the prototype. Then, the investigator asked the participant to sit at the form-filling table (see fig. \ref{fig:insync:setup:plan}) and complete the \ac{TPA} questionnaire and then the demographic data. Subsequently, we continued with the trust game. We handed out the compensation of roughly \trustGameMoney{} as five coins in local currency to the participants. We informed them about the trust game by reading the following instructions: \emph{``Here is your compensation for participating in the study. You have the option to play with the installation. You can give the installation any part of the compensation, and the installation will decide whether to multiply your compensation. It can up to triple the given amount, but it can also choose to keep the entire amount''}. And told them that their participation was completely voluntary. If the participant decided to participate in the game, the investigator opened the coin acceptor cover (see fig. \ref{fig:insync:prototype}B-D) and read the following message: \emph{``The installation accepts coins here. After inserting the coins, please wait 10 seconds for the installation's decision''}.  Participants had as much time as they wanted to decide how much of their compensation they wanted to wager. The investigator stayed in the room for potential assistance. The prototype every time paid back one more coin than the participant inserted. If the participant did not put in any coin or decided not to play, the researcher equalized the participant’s compensation so that each participant ended the study with the same compensation. Finally, the investigator was open to collecting voluntary feedback or comments from participants to use them as inspiration for future work or improvements.

\subsection{Hygiene Measures}

All participants and the investigator were vaccinated against COVID-19 and tested negative using an antigen test on the same day. We ensured that only the investigator and the participant were present in the room. Both the investigator and the participants wore medical face masks throughout the experiment. We disinfected the experimental setup between participants, and all surfaces touched and ventilated the room for 30 minutes. 

\subsection{Analysis}

For the non-parametric analysis of the recorded data, we used the Kruskal–Wallis 1-way analysis of variance with Dunn's tests for multiple comparisons for post-hoc comparisons, correct with Bonferroni's method. We further report the eta-squared \etasquared{} as an estimate of the effect size, classified using Cohen's suggestions as small ($>.0099$), medium ($>.0588$), or large ($>.1379$)~\cite{Cohen1988}. For the analysis of the \dvMoney{}, we employed Shapiro-Wilk’s test and Bartlett's test to check the data for violations of the assumptions of normality and homogeneity of variances, respectively. As the test indicated that the assumption of normality was violated, we continued with a non-parametric analysis as described above.

\subsection{Participants}

We recruited a total of 51 participants (29 identified as female, 22 as male) from our university's mailing list. The participants were aged between 19 and 51 ($\mu = 23.5$, $\sigma = 5.4$). We divided the participants into the three experimental conditions in such a way that they were roughly equally distributed with respect to age and gender, resulting in 17 participants per condition. The participants received around \trustGameMoney{} in the local currency as compensation, which they could use in the trust game as part of the study.

%% file: img/study_setup.tex
\begin{figure*}[t!]
 \centering
 \begin{subfigure}[b]{0.49\textwidth}
    \centering
    \includegraphics[width=\textwidth]{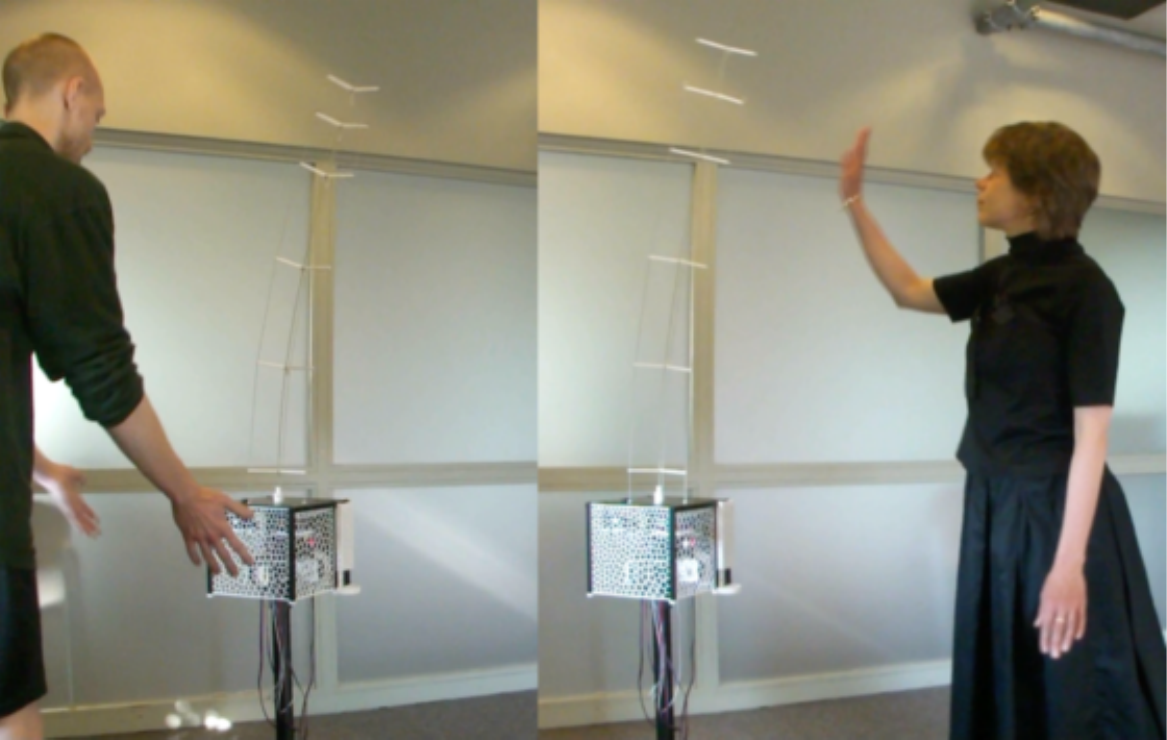}
    \caption{}
    \label{fig:insync:setup:foto}
\end{subfigure}
\hfill
\begin{subfigure}[b]{0.49\textwidth}
    \centering
    \includegraphics[width=\textwidth]{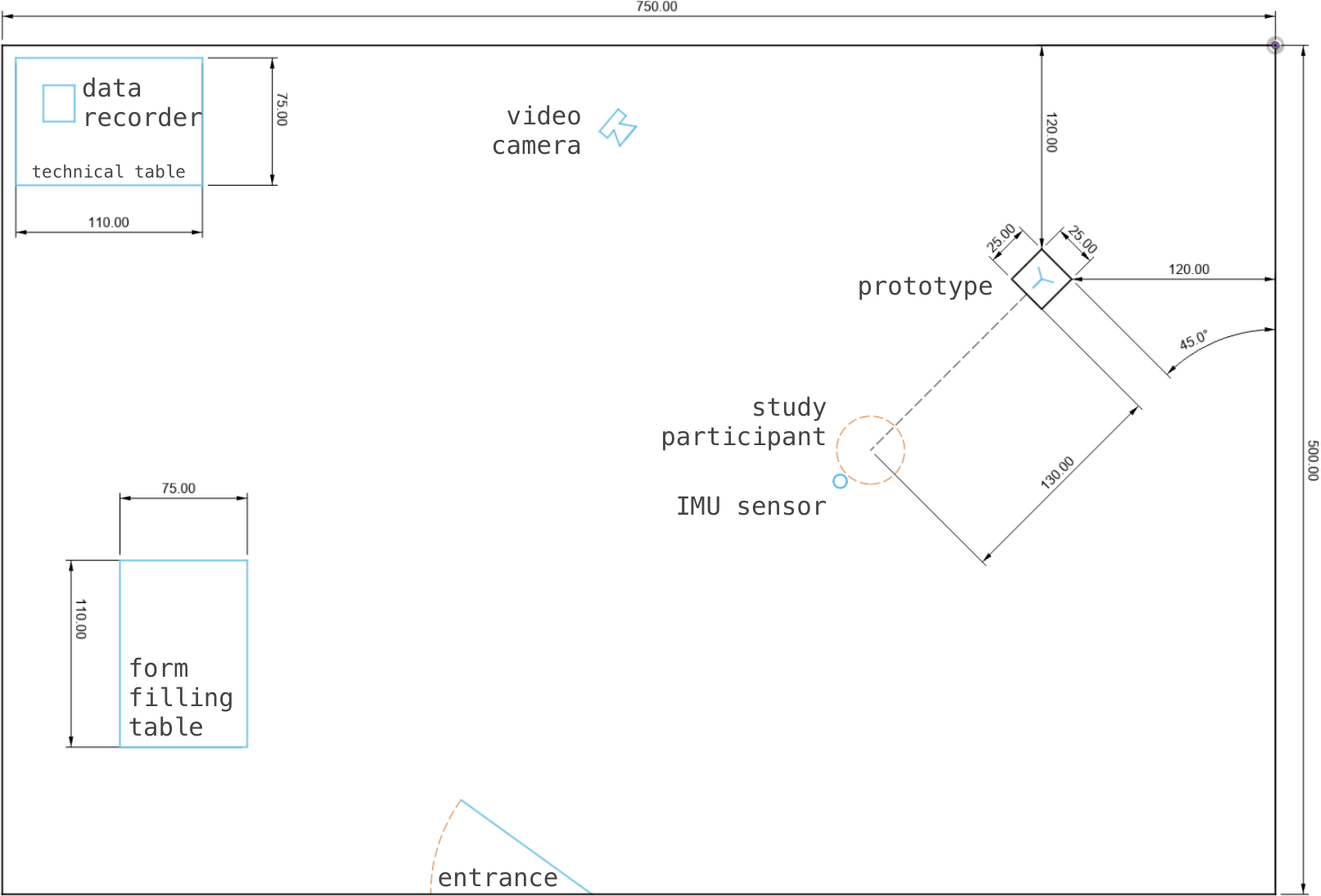}
    \caption{}
    \label{fig:insync:setup:plan}
\end{subfigure}
\caption{The study setup details. (a) Two frames from a video camera showing participants during the free exploration phase. (b) Floor plan of the study setup with labels and dimensions in centimeters.}
\label{fig:insync:methodology:studysetup}
\Description{A two-part figure of the study setup details. (a) Two frames from a video camera showing participants during the free exploration phase. (b) Floor plan of the study setup with labels and dimensions in centimeters.}
\end{figure*}

%% file: content/results.tex
\section{Results}
\label{sec:results}

\input{img/results_sts_money.tex}

In the following section, we report the results of the controlled experiment as described above.

\subsection{\acl{TPA}}

We evaluated the participant's trust in the system using the \ac{TPA} questionnaire.  This 12-item set of Likert scales includes a variety of items sampling trust but also distrust, such as perception of the automation’s deceptive nature or the likelihood of harmful outcomes if it is used. Items that sample distrust must be reverse-coded if used to create a singular trust score \citet{kohn2021measurement}. We reverse-coded items 1 to 5 as they address the distrust dimension. The formula used for a single trust dimension is as follows: \(t=\frac{\sum_{i=1}^5 (8-I_i) + \sum_{i=6}^{12} I_i}{12}\).

We found trust ratings on the \ac{TPA} ranging from \val{3.59}{.74} (\factorMoveLvlRandom{}) over \val{3.90}{.91} (\factorMoveLvlSimple{}) to \val{4.62}{.80} (\factorMoveLvlSynchronized{}), see fig. \ref{fig:insync:results:sts}. A Kruskal-Wallis test indicated a significant (\kruskalwallis{2}{11.18}{<.01}) influence of the \ivMovement{} on the perceived trust with a \efETAsquared{0.19} effect size. Dunn's post-hoc test corrected for multiple comparisons using the Bonferroni method confirmed significantly higher trust ratings for \factorMoveLvlSynchronized{} compared to both \factorMoveLvlSimple{} (\ztest{-2.48}{<.05}) and \factorMoveLvlRandom{} (\ztest{-3.18}{<.01}). We could not find any significant differences between \factorMoveLvlSimple{} and \factorMoveLvlRandom{} (\ztest{.71}{>.05}). 

\input{img/likerttable.tex}

To gain further insights into the participant's attitudes toward the \factorMove{}, we analyzed the individual subscales of the \ac{TPA}. For six subscales, a Kruskal-Wallis test indicated significant differences (see table \ref{tab:insync:results:sts}). Post-hoc tests confirmed significant differences between the \factorMoveLvlSynchronized{} and \factorMoveLvlRandom{} conditions for five subscales. Additionally, we found significant differences between \factorMoveLvlSynchronized{} and \factorMoveLvlSimple{} for two subscales. We could not find significant differences between \factorMoveLvlSimple{} and \factorMoveLvlRandom{} for any subscales. Table~\ref{tab:insync:results:sts} lists the test result for the individual subscales. Further, figure \ref{fig:insync:results:likert} provides a breakdown of the internal distribution of the measured variables for all subscales with significant differences.

\input{img/likert.tex}

\subsection{The Trust Game}

As an additional measurement of the participants' trust toward the prototype, we adapted a method of the trust game as described in section \ref{sec:methodology:design}. All but one participant participated in the game. We found the highest number of inserted coins for the \factorMoveLvlSynchronized{} condition (\val{2.82}{1.67}), followed by \factorMoveLvlRandom{} (\val{2.47}{1.66}) and \factorMoveLvlSimple{} (\val{2.41}{1.33}), see fig. \ref{fig:insync:results:money}. As Shapiro-Wilk's test indicated a violation of the assumption of normality of the residuals that could not be resolved by transforming the data on the log scale, we continued with a non-parametric analysis. However, a subsequent Kruskal-Wallis test did not reveal a significant (\kruskalwallis{2}{0.85}{>.05}) influence of the \factorMove{} of the system on the number of coins inserted.

Further, we analyzed the correlation between the \dvTPA{} and the \dvMoney{} grouped by the \ivMovement{}. We found no significant correlations over all groups. While we could not find a trend for \factorMoveLvlSimple{} (\pearson{.005}{>.05}) and \factorMoveLvlRandom{} (\pearson{-.023}{>.05}) movements, \factorMoveLvlSynchronized{} movements indicated a non-significant trend toward a positive relationship (\pearson{.211}{>.05}). Figure \ref{fig:insync:results:correlation} depicts the pairs and fitted correlation lines.





%% file: img/results_sts_money.tex

\begin{figure*}[t!]
	\centering
	\includegraphics[width=\textwidth]{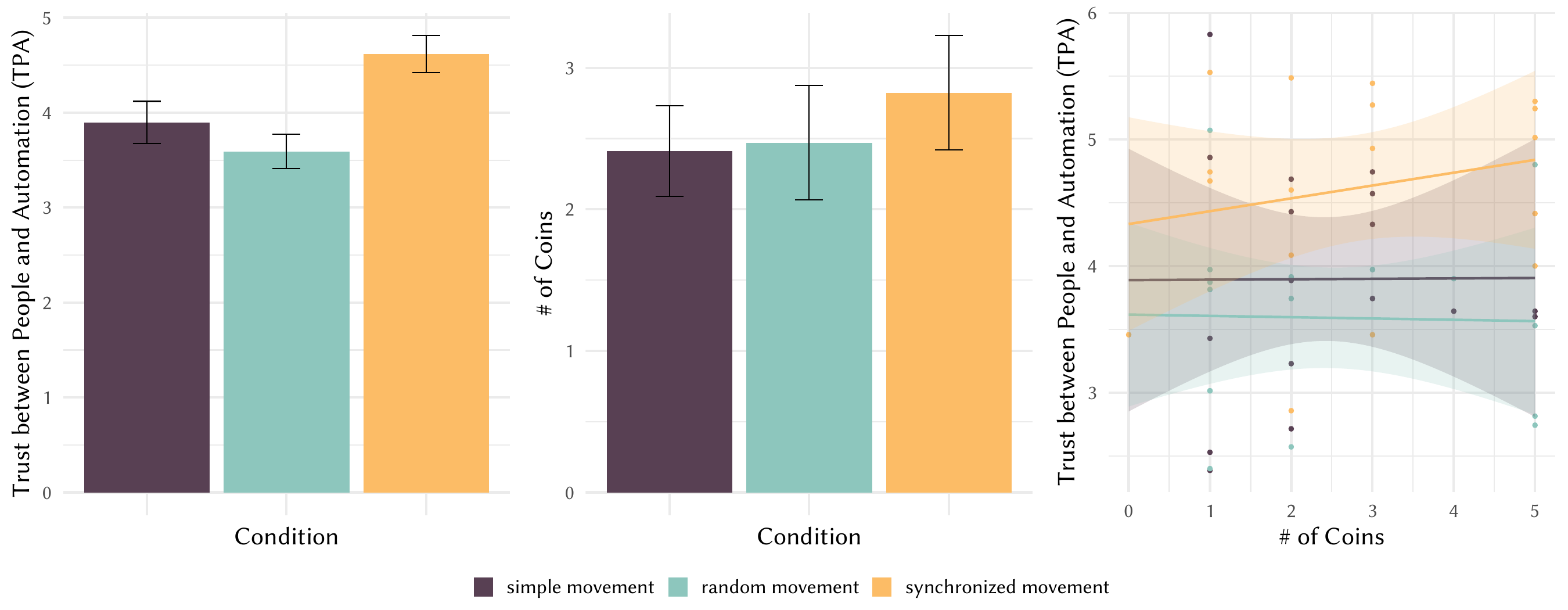}\hfill
	\vspace{-1em}
	\begin{minipage}[t]{.33\linewidth}
		\centering
		\subcaption{\acl{TPA}}\label{fig:insync:results:sts}
	\end{minipage}%
	\begin{minipage}[t]{.33\linewidth}
		\centering
		\subcaption{Money Task}\label{fig:insync:results:money}
\end{minipage}%
	\begin{minipage}[t]{.33\linewidth}
		\centering
		\subcaption{Correlation between \textsc{tpa} and Coin Task}\label{fig:insync:results:correlation}
\end{minipage}%

	\caption{The aggregated results of the (a) \acl{TPA} checklist and (b) the mean number of coins as measured in our experiment. (c) depicts the correlation between the two measurements for the three experimental groups. All error bars depict the standard error.}
	\Description{A three-part figure with charts of the aggregated results of the (a) Trust between People and Automation checklist and (b) the mean number of coins as measured in our experiment. (c) depicts the correlation between the two measurements for the three experimental groups. All error bars depict the standard error.}
	
\end{figure*}

%% file: img/likerttable.tex
\begin{table*}
	\resizebox{\textwidth}{!}{%
		\begin{tabular}{llllllllllccc}
			\hline
			\multicolumn{1}{c}{\multirow{3}{*}{\textbf{Question}}}                                                   & \multicolumn{2}{c}{\textbf{simp}}                       & \multicolumn{2}{c}{\textbf{rand}}                       & \multicolumn{2}{c}{\textbf{sync}}                       & \multicolumn{3}{c}{\textbf{Kruskal-Wallis}}                                          & \multicolumn{3}{c}{\textbf{Dunn's Test}}                                       \\
			\multicolumn{1}{c}{}                                                                                     & \multirow{2}{*}{$\widetilde{x}$} & \multirow{2}{*}{IQR} & \multirow{2}{*}{$\widetilde{x}$} & \multirow{2}{*}{IQR} & \multirow{2}{*}{$\widetilde{x}$} & \multirow{2}{*}{IQR} & \multirow{2}{*}{$\chi^2(2)$} & \multirow{2}{*}{$p$} & \multirow{2}{*}{$\eta_{}^{2}$} & \multicolumn{1}{l}{simp} & \multicolumn{1}{l}{simp} & \multicolumn{1}{l}{rand} \\
			\multicolumn{1}{c}{}                                                                                     &                                  &                      &                                  &                      &                                  &                      &                              &                      &                                & rand                     & sync                     & sync                     \\ \hline
			The system is deceptive.                                                                                 & 2                                & 3                    & 4                                & 4                    & 3                                & 3                    & .74                          & \textgreater{}.05    &                                & \textbf{}                & \textbf{}                & \textbf{}                \\
			\begin{tabular}[c]{@{}l@{}}The system behaves in an\\ underhanded manner.\end{tabular}                   & 3                                & 3                    & 5                                & 3                    & 2                                & 2                    & 7.67                         & \textless{}.05       & .12                            & \textbf{}                & \textbf{}                & *                        \\
			\begin{tabular}[c]{@{}l@{}}I am suspicious of the system's\\ intent, actions or outputs.\end{tabular}    & 3                                & 4                    & 4                                & 2                    & 4                                & 4                    & 1.05                         & \textgreater{}.05    &                                & \textbf{}                & \textbf{}                & \textbf{}                \\
			I am wary of the system.                                                                                 & 5                                & 1                    & 5                                & 2                    & 3                                & 1                    & 6.67                         & \textless{}.05       & .10                            & \textbf{}                & \textbf{}                & \textbf{}                \\
			\begin{tabular}[c]{@{}l@{}}The system's actions will have\\ a harmful or injurious outcome.\end{tabular} & 1                                & 1                    & 1                                & 1                    & 2                                & 2                    & 2.07                         & \textgreater{}.05    &                                & \textbf{}                & \textbf{}                & \textbf{}                \\
			I am confident in the system.                                                                            & 4                                & 1                    & 3                                & 2                    & 4                                & 0                    & 3.73                         & \textgreater{}.05    &                                & \textbf{}                & \textbf{}                & \textbf{}                \\
			The system provides security.                                                                            & 5                                & 3                    & 5                                & 2                    & 5                                & 2                    & .72                          & \textgreater{}.05    &                                & \textbf{}                & \textbf{}                & \textbf{}                \\
			The system has integrity.                                                                                & 4                                & 1                    & 3                                & 2                    & 4                                & 1                    & 11.78                        & \textless{}.01       & .20                            &                          & *                        & **                       \\
			The system is dependable.                                                                                & 3                                & 2                    & 2                                & 2                    & 4                                & 1                    & 10.37                        & \textless{}.01       & .17                            & \textbf{}                & \textbf{}                & **                       \\
			The system is reliable.                                                                                  & 3                                & 2                    & 2                                & 1                    & 3                                & 2                    & 4.03                         & \textgreater{}.05    &                                & \textbf{}                & \textbf{}                & \textbf{}                \\
			I can trust the system.                                                                                  & 3                                & 2                    & 3                                & 2                    & 4                                & 2                    & 9.18                         & \textless{}.05       & .15                            & \textbf{}                & \textbf{}                & *                        \\
			I am familiar with the system.                                                                           & 1                                & 1                    & 1                                & 1                    & 5                                & 2                    & 27.10                        & \textless{}.001      & .52                            & \textbf{}                & ***                      & ***                      \\ \hline
		\end{tabular}%
	}
	\caption{The participant's answers to the individual subscales of the \acl{TPA}. Asterisks refer to the assumed significance levels $p<.05$ (*), $p<.01$ (**) and $p<.001$ (***).}
\label{tab:insync:results:sts}
\end{table*}

%% file: img/likert.tex
\begin{figure*}[ht!]
 \centering
\includegraphics[width=\textwidth]{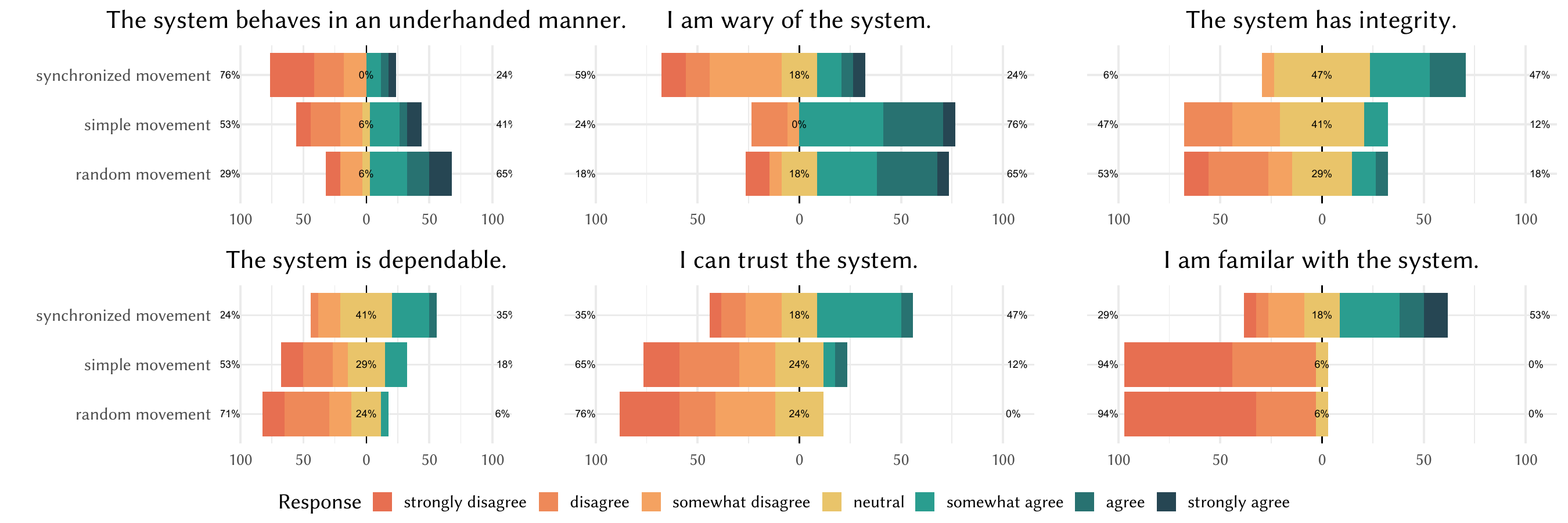}
\caption{Participants' answers to the \acf{TPA} questions. The figure depicts the six statements that provoked significant differences between the three groups.}
\label{fig:insync:results:likert}
\Description{A six-part figure showing participants’ answers to the trust between People and Automation (TPA) questions. The figure depicts the six statements that provoked significant differences between the three groups.}
\end{figure*}

%% file: content/discussion.tex
\section{Discussion}
\label{sec:discussion}

The results of our controlled experiment suggest that movement synchronization can positively influence the feeling of trust of humans toward an \simNonHumRobot{}. In the following section, we discuss our results with regard to the research questions presented above.



\subsection{Synchronized Movement Increases Perceived Trust}

We found that \factorMoveLvlSynchronized{} movements resulted in significantly higher trust ratings on the widely used \acl{TPA} questionnaire compared to both other types of movement, \factorMoveLvlRandom{} (Brownian-like) and \factorMoveLvlSimple{} (sinusoidal waving sideways in one plane). We could not find any significant differences between \factorMoveLvlSimple{} and \factorMoveLvlRandom{} conditions. We attribute similar performance to them belonging to the same underlying category, i.e., unsynchronized movements. To gain a deeper understanding, we analyzed how individual subscales of the \dvTPA{} questionnaire contributed to the result. We found significantly higher ratings for \factorMoveLvlSynchronized{} movements on the perceived dependability, integrity, and familiarity, as well as lower ratings for the wariness and expected underhanded behavior of the prototype. We hypothesize that this is caused by participants feeling bonded with the robot through synchronisation and perceiving a smoother interaction~\cite{stel2010mimicry}.

Our results demonstrate a distinct influence of the motion patterns on the perceived trust toward the prototype. We collected these results using an abstract prototype to rule out the influence of a possible anthropomorphization, which has been suggested in previous work \cite{Salem2015, Langer2019, deVisser2017}. Thus, we reject the null hypothesis (i.e., the absence of an influence of synchronized motion on perceived trust). These results are in line with previous work investigating the influence of motion synchronization between people on perceived trust (see section \ref{sec:relatedwork}).

Therefore, in this paper, we have demonstrated a new strategy based on synchronization to establish a trusting relationship between humans and robots. In contrast to solutions proposed by previous work \cite{calvo2020effects, naneva2020systematic, salem2015would, christoforakos2021can}, our approach does not depend on the robot's external form or complex behavior. It is thus also suitable for \simNonHumRobots{}. We consider our research to present a significant contribution as, to the best of our knowledge, there is no prior work exploring synchronization as a way to increase trust between humans and \simNonHumRobots{}.

\subsection{Synchronized Movement Does Not Affect the Willingness to Spend Money in Our Version of the Trust Game}

We found the highest number of coins inserted for the \factorMoveLvlSynchronized{} condition. Still, we could not find a significant influence of the \factorMove{} of the prototype on the number of coins inserted. Also, we found no significant correlations between the \dvTPA{} and the \dvMoney{}.

We attribute this lack of a significant effect on the \dvMoney{}  to several factors. First, our observations and informal interviews showed that people treated the game as gambling. We assume that this is explained by the fact that the amount of compensation for participation was not high enough. Thus, the risk was low, so participants did not differentiate the amounts they wagered on the game. This is in line with findings from prior work indicating that people's averse to gambling can influence the results of the trust game~\cite{chetty2021trust}.

Second, we speculate that the time between interacting with the prototype and the trust game was too long because participants had to complete the \dvTPA{} in between. During the research procedure design process, we considered making the trust game a part of the interaction phase. We dismissed this idea later, however, as the gambling results would potentially override the subtle effect of synchronization on trust and influence the result of \dvTPA{}. As a result, participants would assess not so much trust as satisfaction with the number of coins returned by the robot. Further, from our observation, we conclude that the participants were surprised by the proposal to insert coins into the prototype. The situation was strange and incomprehensible to them. Finally, the investigator was present in the room, so we may have measured confidence toward the investigator rather than the prototype.

In future studies, we recommend leaving the participant alone with the prototype during this study phase. We also recommend conducting the trust game directly after interacting with the prototype or making it a part of the free interaction phase if the trust game is the only measure.


%% file: content/limitations.tex
\section{Limitations and Future Work}
\label{sec:limitions}

While we are convinced that our results present a viable contribution to the body of research on trust in human-robot interaction, the study design, as well as the results, impose limitations and directions for future work. In the following section, we discuss them.

\subsection{External Validity and Real-World Applicability}

We deliberately decided to use a form as abstract as possible in our experiment. We opted for this approach to avoid anthropomorphizing the robot’s form and associations with other animate forms carrying specific attitudes or emotions. We also avoided association with all kinds of popular robots. We are confident that the results enable a broad range of real-world application areas because everywhere where the form or complex behavior of the robot is limited, we can use dynamic, in this case, synchronization, as a means to design a trustworthy robot. However, it is an open question on how the results translate to other robot forms. For example, a robotic arm associated with an industrial robot will not eliminate the synchronization effect. Future work is necessary to conclude these challenges.

Further, while reviewing video recordings, we found that in the \factorMoveLvlSynchronized{} condition, at the moment when the participant discovered that the robot synchronizes with their movements, most of them had a strong positive reaction (broad and long smile) and started exploring the prototype more actively. But over time, there was boredom with the lack of new prototype behavior. This is in line with the works of \citet{ravreby2022liking}, who found that novelty or more complex behavior, even at the cost of losing or having more difficulty in achieving synchronization, produces much better results than keeping synchronization. It would be interesting to investigate whether this effect can be reproduced in interaction with the \simNonHumRobot{}.

Considering a broad body of works in HCI on human interactions mediated by technology, especially those through physical objects \cite{brave1998tangible}, it is an intriguing open research question of how our prototype would be perceived in such interactions. The basic idea is that the participant's moves will be mimicked by a remote prototype in front of another person and vice versa. In such a setup, we could, for example, manipulate the coupling parameter of synchronization of two humans remotely interacting, looking at their perception of the relation between them \cite{vallacher2005dynamics}.



\subsection{Synchronization during Collaborative Tasks}

Besides the abstract body of our prototype, we further opted to build a system that basically serves no purpose other than being there and moving with the participant. We chose this path to provide a solid baseline and exclude influencing factors from an actual task of the system. Therefore, we analyzed synchronization as a tool for shaping the attitude toward the robot. However, it is still an open challenge how our results apply to collaborative scenarios where we as humans are working together with a system in a collaborative task. 

Here we see interesting questions in many areas. For example, for speakers with voice assistants, synchronization could increase the confidence in the assistant by giving them significant physical properties through the dynamics of movement. In such a scenario, the assistant synchronizes with the dynamics of human movements to increase affective trust. Compared to prior work, the goal is not to provide fitting gesticulation in harmony with her voice but to establish synchronization with the user. Further, with autonomous cars on the horizon, synchronization of movements of visual and physical elements of the car cockpit and other interfaces could help establish a relationship of trust, increasing the sense of comfort with unfamiliar situations and understanding the intentions of the car. 

Further, we did not enforce or restrict any movements of the participants and, thus, deliberately left it up to the participants how and if they wanted to interact with the system. We chose this approach to avoid biasing the provided baseline in any direction by external influences and to recreate a situation in which people interact freely with a system without a specific goal. While we did not collect the movement quantities, we hypothesise, based on our observations, that the amount of movement could have an influence on the perception of trust and - vice versa - the strength of trust could have an influence on the movement of the participants.

Lastly, in some safety-critical areas, it could also be valuable to decrease trust. It may be interesting to research if the user unconsciously wants to do potentially harmful actions to himself or the machine. Then breaking synchronization can signal the user to focus attention and revise the action and, as a further consequence, decrease trust in a potentially dangerous machine. Future work is needed to conclude these challenges.



\subsection{Synchronicity Beyond Upper-body Movements}

We explored synchronization only with upper-body movements. We opted for this approach to explore the simplest possible patterns leading to synchronization at the level of body movement.

Thus, it remains an open question how other body movements, such as head nodding, gesturing with the hands, or leg movements, can be mapped to induce synchronization. Beyond body movements, there are different body signals like breathing rate or heartbeat, which could also be used to establish synchronization.

On the robot side, we have only investigated a direct mapping of the participant’s movement, i.e., a chameleon-like rendering. It is an additional open question whether we can transfer synchronization patterns to other output modalities by keeping their dynamics, for example, mapping body movements to the brightness of a lamp. Further work is needed to answer these questions.

\subsection{Trustworthy Robot and Ethics}

In HRI, trust plays a crucial role and is strongly linked to persuasiveness. Under-trust or over-trust may have severe or even dangerous consequences~\cite{Salem2015}. Our results indicate that the synchronization of \simNonHumRobots{} to the movement patterns of human users can induce a sense of trust in people. These results can, on the one hand, provide the basis for developing systems that use this finding to more easily and quickly establish a relationship of trust with users. On the other hand, however, like different approaches to gaining and maintaining trust in robots, such as anthropomorphization, this effect can be exploited by malicious actors and lead people to trust untrustworthy entities. We consider the investigation of ethical aspects and assessing the possible dangers a vital field of research. Our work, by highlighting the role of synchronization, provides useful starting points for future work in these fields.

As a possible solution, certification and regulation by independent entities~\cite{shneiderman2020bridging} could increase the trustworthiness of interacting with such systems. Even though synchronization can influence the relationship with the machine on an affective level (e.g., feelings, emotions), such regulation could help to ensure that robots are regarded more as tools that we use to improve our own skills and accelerate progress along our own paths~\cite{bryson2010robots}.

%% file: content/conclusion.tex
\section{Conclusion}
\label{sec:conclusion}

\balance

In this paper, we explored movement synchronization as a dynamical approach that can be applied in the design of \simNonHumRobots{} to increase trust. We contributed by design and implementation of a prototype system and the results of a controlled experiment with 51 participants exploring our concept in a between-subjects design. We found significantly higher ratings on trust between people and automation in an established questionnaire. However, we could not find an influence on the willingness to spend money in a trust game. Taken together, our results strongly suggest a positive effect of synchronized movement on the participants' feeling of trust toward an \simNonHumRobot{}. The presented result is also important for designers because it points to dynamics as a design strategy and shows synchronization as a tool for designing trustworthy \simNonHumRobots{}.

%% file: meta/acks.tex
\begin{acks}
	We thank our anonymous reviewers for their valuable comments and suggestions and all participants who took part in our experiment. This work has been funded by the \grantsponsor{h2020}{European Union's Horizon 2020 research and innovation program}{http://ec.europa.eu/programmes/horizon2020/en} under grant agreement No.	\grantnum[https://www.humane-ai.eu/]{h2020}{952026}.

\end{acks}

%% file: insync.bbl

\begin{thebibliography}{93}


\ifx \showCODEN    \undefined \def \showCODEN     #1{\unskip}     \fi
\ifx \showDOI      \undefined \def \showDOI       #1{#1}\fi
\ifx \showISBNx    \undefined \def \showISBNx     #1{\unskip}     \fi
\ifx \showISBNxiii \undefined \def \showISBNxiii  #1{\unskip}     \fi
\ifx \showISSN     \undefined \def \showISSN      #1{\unskip}     \fi
\ifx \showLCCN     \undefined \def \showLCCN      #1{\unskip}     \fi
\ifx \shownote     \undefined \def \shownote      #1{#1}          \fi
\ifx \showarticletitle \undefined \def \showarticletitle #1{#1}   \fi
\ifx \showURL      \undefined \def \showURL       {\relax}        \fi
\providecommand\bibfield[2]{#2}
\providecommand\bibinfo[2]{#2}
\providecommand\natexlab[1]{#1}
\providecommand\showeprint[2][]{arXiv:#2}

\bibitem[\protect\citeauthoryear{Aaltonen, Arvola, Heikkil{\"a}, and
  Lammi}{Aaltonen et~al\mbox{.}}{2017}]%
        {Aaltonen2017}
\bibfield{author}{\bibinfo{person}{Iina Aaltonen}, \bibinfo{person}{Anne
  Arvola}, \bibinfo{person}{P{\"a}ivi Heikkil{\"a}}, {and}
  \bibinfo{person}{Hanna Lammi}.} \bibinfo{year}{2017}\natexlab{}.
\newblock \showarticletitle{Hello {{Pepper}}, {{May I Tickle You}}?
  {{Children}}'s and {{Adults}}' {{Responses}} to an {{Entertainment Robot}} at
  a {{Shopping Mall}}}. In \bibinfo{booktitle}{\emph{Proceedings of the
  {{Companion}} of the 2017 {{ACM}}/{{IEEE International Conference}} on
  {{Human-Robot Interaction}}}} \emph{(\bibinfo{series}{{{HRI}} '17})}.
  \bibinfo{publisher}{{Association for Computing Machinery}},
  \bibinfo{address}{{New York, NY, USA}}, \bibinfo{pages}{53--54}.
\newblock
\showISBNx{978-1-4503-4885-0}
\urldef\tempurl%
\url{https://doi.org/10.1145/3029798.3038362}
\showDOI{\tempurl}


\bibitem[\protect\citeauthoryear{Alarcon, Capiola, Hamdan, Lee, and
  Jessup}{Alarcon et~al\mbox{.}}{2023}]%
        {ALARCON2023103858}
\bibfield{author}{\bibinfo{person}{Gene~M. Alarcon}, \bibinfo{person}{August
  Capiola}, \bibinfo{person}{Izz~Aldin Hamdan}, \bibinfo{person}{Michael~A.
  Lee}, {and} \bibinfo{person}{Sarah~A. Jessup}.}
  \bibinfo{year}{2023}\natexlab{}.
\newblock \showarticletitle{Differential Biases in Human-Human versus
  Human-Robot Interactions}.
\newblock \bibinfo{journal}{\emph{Applied Ergonomics}}  \bibinfo{volume}{106}
  (\bibinfo{date}{Jan.} \bibinfo{year}{2023}), \bibinfo{pages}{103858}.
\newblock
\showISSN{0003-6870}
\urldef\tempurl%
\url{https://doi.org/10.1016/j.apergo.2022.103858}
\showDOI{\tempurl}


\bibitem[\protect\citeauthoryear{Baker, Phillips, Ullman, and Keebler}{Baker
  et~al\mbox{.}}{2018}]%
        {baker2018toward}
\bibfield{author}{\bibinfo{person}{Anthony~L. Baker},
  \bibinfo{person}{Elizabeth~K. Phillips}, \bibinfo{person}{Daniel Ullman},
  {and} \bibinfo{person}{Joseph~R. Keebler}.} \bibinfo{year}{2018}\natexlab{}.
\newblock \showarticletitle{Toward an {{Understanding}} of {{Trust Repair}} in
  {{Human-Robot Interaction}}: {{Current Research}} and {{Future Directions}}}.
\newblock \bibinfo{journal}{\emph{ACM Transactions on Interactive Intelligent
  Systems}} \bibinfo{volume}{8}, \bibinfo{number}{4} (\bibinfo{date}{Nov.}
  \bibinfo{year}{2018}), \bibinfo{pages}{30:1--30:30}.
\newblock
\showISSN{2160-6455}
\urldef\tempurl%
\url{https://doi.org/10.1145/3181671}
\showDOI{\tempurl}


\bibitem[\protect\citeauthoryear{Baron, Amazeen, and Beek}{Baron
  et~al\mbox{.}}{1994}]%
        {baron1994local}
\bibfield{author}{\bibinfo{person}{Reuben~M. Baron},
  \bibinfo{person}{Polemnia~G. Amazeen}, {and} \bibinfo{person}{Peter~J.
  Beek}.} \bibinfo{year}{1994}\natexlab{}.
\newblock \showarticletitle{Local and Global Dynamics of Social Relations.}
\newblock In \bibinfo{booktitle}{\emph{Dynamical Systems in Social
  Psychology.}} \bibinfo{publisher}{{Academic Press}}, \bibinfo{address}{{San
  Diego, CA, US}}, \bibinfo{pages}{111--138}.
\newblock
\showISBNx{0-12-709990-5 (Hardcover)}


\bibitem[\protect\citeauthoryear{Berg, Dickhaut, and McCabe}{Berg
  et~al\mbox{.}}{1995}]%
        {BERG1995122}
\bibfield{author}{\bibinfo{person}{Joyce Berg}, \bibinfo{person}{John
  Dickhaut}, {and} \bibinfo{person}{Kevin McCabe}.}
  \bibinfo{year}{1995}\natexlab{}.
\newblock \showarticletitle{Trust, Reciprocity, and Social History}.
\newblock \bibinfo{journal}{\emph{Games and Economic Behavior}}
  \bibinfo{volume}{10}, \bibinfo{number}{1} (\bibinfo{year}{1995}),
  \bibinfo{pages}{122--142}.
\newblock
\showISSN{0899-8256}
\urldef\tempurl%
\url{https://doi.org/10.1006/game.1995.1027}
\showDOI{\tempurl}


\bibitem[\protect\citeauthoryear{Berx, Adriaensen, Decr{\'e}, and
  Pintelon}{Berx et~al\mbox{.}}{2022}]%
        {safety8030048}
\bibfield{author}{\bibinfo{person}{Nicole Berx}, \bibinfo{person}{Arie
  Adriaensen}, \bibinfo{person}{Wilm Decr{\'e}}, {and} \bibinfo{person}{Liliane
  Pintelon}.} \bibinfo{year}{2022}\natexlab{}.
\newblock \showarticletitle{Assessing {{System-Wide Safety Readiness}} for
  {{Successful Human}}\textendash{{Robot Collaboration Adoption}}}.
\newblock \bibinfo{journal}{\emph{Safety}} \bibinfo{volume}{8},
  \bibinfo{number}{3} (\bibinfo{date}{Sept.} \bibinfo{year}{2022}),
  \bibinfo{pages}{48}.
\newblock
\showISSN{2313-576X}
\urldef\tempurl%
\url{https://doi.org/10.3390/safety8030048}
\showDOI{\tempurl}


\bibitem[\protect\citeauthoryear{Bos, Olson, Gergle, Olson, and Wright}{Bos
  et~al\mbox{.}}{2002}]%
        {bos2002effects}
\bibfield{author}{\bibinfo{person}{Nathan Bos}, \bibinfo{person}{Judy Olson},
  \bibinfo{person}{Darren Gergle}, \bibinfo{person}{Gary Olson}, {and}
  \bibinfo{person}{Zach Wright}.} \bibinfo{year}{2002}\natexlab{}.
\newblock \showarticletitle{Effects of Four Computer-Mediated Communications
  Channels on Trust Development}. In \bibinfo{booktitle}{\emph{Proceedings of
  the {{SIGCHI Conference}} on {{Human Factors}} in {{Computing Systems}}}}
  \emph{(\bibinfo{series}{{{CHI}} '02})}. \bibinfo{publisher}{{Association for
  Computing Machinery}}, \bibinfo{address}{{New York, NY, USA}},
  \bibinfo{pages}{135--140}.
\newblock
\showISBNx{978-1-58113-453-7}
\urldef\tempurl%
\url{https://doi.org/10.1145/503376.503401}
\showDOI{\tempurl}


\bibitem[\protect\citeauthoryear{Brave, Ishii, and Dahley}{Brave
  et~al\mbox{.}}{1998}]%
        {brave1998tangible}
\bibfield{author}{\bibinfo{person}{Scott Brave}, \bibinfo{person}{Hiroshi
  Ishii}, {and} \bibinfo{person}{Andrew Dahley}.}
  \bibinfo{year}{1998}\natexlab{}.
\newblock \showarticletitle{Tangible Interfaces for Remote Collaboration and
  Communication}. In \bibinfo{booktitle}{\emph{Proceedings of the 1998 {{ACM}}
  Conference on {{Computer}} Supported Cooperative Work}}
  \emph{(\bibinfo{series}{{{CSCW}} '98})}. \bibinfo{publisher}{{Association for
  Computing Machinery}}, \bibinfo{address}{{New York, NY, USA}},
  \bibinfo{pages}{169--178}.
\newblock
\showISBNx{978-1-58113-009-6}
\urldef\tempurl%
\url{https://doi.org/10.1145/289444.289491}
\showDOI{\tempurl}


\bibitem[\protect\citeauthoryear{Brown, Gray, McHardy, and Taylor}{Brown
  et~al\mbox{.}}{2015}]%
        {Brown2015}
\bibfield{author}{\bibinfo{person}{Sarah Brown}, \bibinfo{person}{Daniel Gray},
  \bibinfo{person}{Jolian McHardy}, {and} \bibinfo{person}{Karl Taylor}.}
  \bibinfo{year}{2015}\natexlab{}.
\newblock \showarticletitle{Employee Trust and Workplace Performance}.
\newblock \bibinfo{journal}{\emph{Journal of Economic Behavior \&
  Organization}}  \bibinfo{volume}{116} (\bibinfo{date}{Aug.}
  \bibinfo{year}{2015}), \bibinfo{pages}{361--378}.
\newblock
\showISSN{0167-2681}
\urldef\tempurl%
\url{https://doi.org/10.1016/j.jebo.2015.05.001}
\showDOI{\tempurl}


\bibitem[\protect\citeauthoryear{Bryson and Wilks}{Bryson and Wilks}{2010}]%
        {bryson2010robots}
\bibfield{author}{\bibinfo{person}{Joanna~J. Bryson} {and}
  \bibinfo{person}{Yorick Wilks}.} \bibinfo{year}{2010}\natexlab{}.
\newblock \showarticletitle{Robots Should Be Slaves}.
\newblock In \bibinfo{booktitle}{\emph{Close {{Engagements}} with {{Artificial
  Companions}}}}. \bibinfo{publisher}{{John Benjamins Publishing Company}},
  \bibinfo{address}{{Amsterdam, Noord-holland, 1033, Netherlands}},
  \bibinfo{pages}{63--74}.
\newblock
\showISBNx{978-90-272-8840-0 978-90-272-4994-4}
\urldef\tempurl%
\url{https://doi.org/10.1075/nlp.8.11bry}
\showDOI{\tempurl}


\bibitem[\protect\citeauthoryear{{Calvo-Barajas}, Perugia, and
  Castellano}{{Calvo-Barajas} et~al\mbox{.}}{2020}]%
        {calvo2020effects}
\bibfield{author}{\bibinfo{person}{Natalia {Calvo-Barajas}},
  \bibinfo{person}{Giulia Perugia}, {and} \bibinfo{person}{Ginevra
  Castellano}.} \bibinfo{year}{2020}\natexlab{}.
\newblock \showarticletitle{The {{Effects}} of {{Robot}}'s {{Facial
  Expressions}} on {{Children}}'s {{First Impressions}} of
  {{Trustworthiness}}}. In \bibinfo{booktitle}{\emph{2020 29th {{IEEE
  International Conference}} on {{Robot}} and {{Human Interactive
  Communication}} ({{RO-MAN}})}}. \bibinfo{publisher}{{IEEE Comput. Soc}},
  \bibinfo{address}{{New York City, United States}}, \bibinfo{pages}{165--171}.
\newblock
\showISSN{1944-9437}
\urldef\tempurl%
\url{https://doi.org/10.1109/RO-MAN47096.2020.9223456}
\showDOI{\tempurl}


\bibitem[\protect\citeauthoryear{Chartrand and Bargh}{Chartrand and
  Bargh}{1999}]%
        {chartrand1999chameleon}
\bibfield{author}{\bibinfo{person}{Tanya~L. Chartrand} {and}
  \bibinfo{person}{John~A. Bargh}.} \bibinfo{year}{1999}\natexlab{}.
\newblock \showarticletitle{The Chameleon Effect: {{The}} Perception\textendash
  Behavior Link and Social Interaction}.
\newblock \bibinfo{journal}{\emph{Journal of Personality and Social
  Psychology}}  \bibinfo{volume}{76} (\bibinfo{year}{1999}),
  \bibinfo{pages}{893--910}.
\newblock
\showISSN{1939-1315}
\urldef\tempurl%
\url{https://doi.org/10.1037/0022-3514.76.6.893}
\showDOI{\tempurl}


\bibitem[\protect\citeauthoryear{Chartrand and {van Baaren}}{Chartrand and {van
  Baaren}}{2009}]%
        {chartrand2009human}
\bibfield{author}{\bibinfo{person}{Tanya~L. Chartrand} {and}
  \bibinfo{person}{Rick {van Baaren}}.} \bibinfo{year}{2009}\natexlab{}.
\newblock \showarticletitle{Human {{Mimicry}}}.
\newblock In \bibinfo{booktitle}{\emph{Advances in {{Experimental Social
  Psychology}}}}. Vol.~\bibinfo{volume}{41}. \bibinfo{publisher}{{Academic
  Press}}, \bibinfo{address}{{Cambridge, Massachusetts, United States}},
  \bibinfo{pages}{219--274}.
\newblock
\urldef\tempurl%
\url{https://doi.org/10.1016/S0065-2601(08)00405-X}
\showDOI{\tempurl}


\bibitem[\protect\citeauthoryear{Cheng, Zhu, Hu, Zhou, Pan, and Hu}{Cheng
  et~al\mbox{.}}{2022}]%
        {Cheng2022}
\bibfield{author}{\bibinfo{person}{Xiaojun Cheng}, \bibinfo{person}{Yujiao
  Zhu}, \bibinfo{person}{Yinying Hu}, \bibinfo{person}{Xiaolin Zhou},
  \bibinfo{person}{Yafeng Pan}, {and} \bibinfo{person}{Yi Hu}.}
  \bibinfo{year}{2022}\natexlab{}.
\newblock \showarticletitle{Integration of Social Status and Trust through
  Interpersonal Brain Synchronization}.
\newblock \bibinfo{journal}{\emph{NeuroImage}}  \bibinfo{volume}{246}
  (\bibinfo{date}{Feb.} \bibinfo{year}{2022}), \bibinfo{pages}{118777}.
\newblock
\showISSN{1053-8119}
\urldef\tempurl%
\url{https://doi.org/10.1016/j.neuroimage.2021.118777}
\showDOI{\tempurl}


\bibitem[\protect\citeauthoryear{Chetty, Hofmeyr, Kincaid, and Monroe}{Chetty
  et~al\mbox{.}}{2021}]%
        {chetty2021trust}
\bibfield{author}{\bibinfo{person}{Rinelle Chetty}, \bibinfo{person}{Andre
  Hofmeyr}, \bibinfo{person}{Harold Kincaid}, {and} \bibinfo{person}{Brian
  Monroe}.} \bibinfo{year}{2021}\natexlab{}.
\newblock \showarticletitle{The {{Trust Game Does Not}} ({{Only}}) {{Measure
  Trust}}: {{The Risk-Trust Confound Revisited}}}.
\newblock \bibinfo{journal}{\emph{Journal of Behavioral and Experimental
  Economics}}  \bibinfo{volume}{90} (\bibinfo{date}{Feb.}
  \bibinfo{year}{2021}), \bibinfo{pages}{101520}.
\newblock
\showISSN{2214-8043}
\urldef\tempurl%
\url{https://doi.org/10.1016/j.socec.2020.101520}
\showDOI{\tempurl}


\bibitem[\protect\citeauthoryear{Christoforakos, Gallucci, {Surmava-Gro{\ss}e},
  Ullrich, and Diefenbach}{Christoforakos et~al\mbox{.}}{2021}]%
        {christoforakos2021can}
\bibfield{author}{\bibinfo{person}{Lara Christoforakos},
  \bibinfo{person}{Alessio Gallucci}, \bibinfo{person}{Tinatini
  {Surmava-Gro{\ss}e}}, \bibinfo{person}{Daniel Ullrich}, {and}
  \bibinfo{person}{Sarah Diefenbach}.} \bibinfo{year}{2021}\natexlab{}.
\newblock \showarticletitle{Can Robots Earn Our Trust the Same Way Humans Do?
  {{A}} Systematic Exploration of Competence, Warmth, and Anthropomorphism as
  Determinants of Trust Development in {{HRI}}}.
\newblock \bibinfo{journal}{\emph{Frontiers in Robotics and AI}}
  \bibinfo{volume}{8} (\bibinfo{year}{2021}).
\newblock
\showISSN{2296-9144}
\urldef\tempurl%
\url{https://doi.org/10.3389/frobt.2021.640444}
\showDOI{\tempurl}


\bibitem[\protect\citeauthoryear{Cohen}{Cohen}{1988}]%
        {Cohen1988}
\bibfield{author}{\bibinfo{person}{Jacob Cohen}.}
  \bibinfo{year}{1988}\natexlab{}.
\newblock \bibinfo{booktitle}{\emph{Statistical {{Power Analysis}} for the
  {{Behavioral Sciences}}} (\bibinfo{edition}{second} ed.)}.
\newblock \bibinfo{publisher}{{Routledge}}, \bibinfo{address}{{New York}}.
\newblock
\showISBNx{978-0-203-77158-7}
\urldef\tempurl%
\url{https://doi.org/10.4324/9780203771587}
\showDOI{\tempurl}


\bibitem[\protect\citeauthoryear{Correia, {Alves-Oliveira}, Maia, Ribeiro,
  Petisca, Melo, and Paiva}{Correia et~al\mbox{.}}{2016}]%
        {correia2016just}
\bibfield{author}{\bibinfo{person}{Filipa Correia},
  \bibinfo{person}{Patr{\'i}cia {Alves-Oliveira}}, \bibinfo{person}{Nuno Maia},
  \bibinfo{person}{Tiago Ribeiro}, \bibinfo{person}{Sofia Petisca},
  \bibinfo{person}{Francisco~S. Melo}, {and} \bibinfo{person}{Ana Paiva}.}
  \bibinfo{year}{2016}\natexlab{}.
\newblock \showarticletitle{Just Follow the Suit! {{Trust}} in Human-Robot
  Interactions during Card Game Playing}. In \bibinfo{booktitle}{\emph{2016
  25th {{IEEE International Symposium}} on {{Robot}} and {{Human Interactive
  Communication}} ({{RO-MAN}})}}. \bibinfo{publisher}{{IEEE Comput. Soc}},
  \bibinfo{address}{{New York, NY, US}}, \bibinfo{pages}{507--512}.
\newblock
\showISSN{1944-9437}
\urldef\tempurl%
\url{https://doi.org/10.1109/ROMAN.2016.7745165}
\showDOI{\tempurl}


\bibitem[\protect\citeauthoryear{{Cox-George} and Bewley}{{Cox-George} and
  Bewley}{2018}]%
        {CoxGeorge2018}
\bibfield{author}{\bibinfo{person}{Chantal {Cox-George}} {and}
  \bibinfo{person}{Susan Bewley}.} \bibinfo{year}{2018}\natexlab{}.
\newblock \showarticletitle{I, {{Sex Robot}}: The Health Implications of the
  Sex Robot Industry}.
\newblock \bibinfo{journal}{\emph{BMJ Sexual \& Reproductive Health}}
  \bibinfo{volume}{44}, \bibinfo{number}{3} (\bibinfo{date}{July}
  \bibinfo{year}{2018}), \bibinfo{pages}{161--164}.
\newblock
\showISSN{2515-1991, 2515-2009}
\urldef\tempurl%
\url{https://doi.org/10.1136/bmjsrh-2017-200012}
\showDOI{\tempurl}


\bibitem[\protect\citeauthoryear{Crick, Munz, and Scassellati}{Crick
  et~al\mbox{.}}{2006}]%
        {crick2006synchronization}
\bibfield{author}{\bibinfo{person}{Christopher Crick}, \bibinfo{person}{Matthew
  Munz}, {and} \bibinfo{person}{Brian Scassellati}.}
  \bibinfo{year}{2006}\natexlab{}.
\newblock \showarticletitle{Synchronization in {{Social Tasks}}: {{Robotic
  Drumming}}}. In \bibinfo{booktitle}{\emph{{{ROMAN}} 2006 - {{The}} 15th
  {{IEEE International Symposium}} on {{Robot}} and {{Human Interactive
  Communication}}}}. \bibinfo{publisher}{{IEEE Comput. Soc}},
  \bibinfo{address}{{New York, NY, US}}, \bibinfo{pages}{97--102}.
\newblock
\showISSN{1944-9437}
\urldef\tempurl%
\url{https://doi.org/10.1109/ROMAN.2006.314401}
\showDOI{\tempurl}


\bibitem[\protect\citeauthoryear{Daudi, Hauge, and Thoben}{Daudi
  et~al\mbox{.}}{2016}]%
        {daudi2016effects}
\bibfield{author}{\bibinfo{person}{Morice Daudi},
  \bibinfo{person}{Jannicke~Baalsrud Hauge}, {and}
  \bibinfo{person}{Klaus-Dieter Thoben}.} \bibinfo{year}{2016}\natexlab{}.
\newblock \showarticletitle{Effects of {{Decision Synchronization}} on
  {{Trust}} in {{Collaborative Networks}}}. In
  \bibinfo{booktitle}{\emph{Collaboration in a {{Hyperconnected World}}}}
  \emph{(\bibinfo{series}{{{IFIP Advances}} in {{Information}} and
  {{Communication Technology}}})}, \bibfield{editor}{\bibinfo{person}{Hamideh
  Afsarmanesh}, \bibinfo{person}{Luis~M. {Camarinha-Matos}}, {and}
  \bibinfo{person}{Ant{\'o}nio Lucas~Soares}} (Eds.).
  \bibinfo{publisher}{{Springer International Publishing}},
  \bibinfo{address}{{Cham}}, \bibinfo{pages}{215--227}.
\newblock
\showISBNx{978-3-319-45390-3}
\urldef\tempurl%
\url{https://doi.org/10.1007/978-3-319-45390-3_19}
\showDOI{\tempurl}


\bibitem[\protect\citeauthoryear{{de Fine Licht} and Br{\"u}lde}{{de Fine
  Licht} and Br{\"u}lde}{2021}]%
        {de2021defining}
\bibfield{author}{\bibinfo{person}{Karl {de Fine Licht}} {and}
  \bibinfo{person}{Bengt Br{\"u}lde}.} \bibinfo{year}{2021}\natexlab{}.
\newblock \showarticletitle{On Defining ``{{Reliance}}'' and ``{{Trust}}'':
  {{Purposes}}, Conditions of Adequacy, and New Definitions}.
\newblock \bibinfo{journal}{\emph{Philosophia}} \bibinfo{volume}{49},
  \bibinfo{number}{5} (\bibinfo{year}{2021}), \bibinfo{pages}{1981--2001}.
\newblock
\urldef\tempurl%
\url{https://doi.org/10.1007/s11406-021-00339-1}
\showDOI{\tempurl}


\bibitem[\protect\citeauthoryear{De~Visser, Monfort, Goodyear, Lu, O'Hara, Lee,
  Parasuraman, and Krueger}{De~Visser et~al\mbox{.}}{2017}]%
        {deVisser2017}
\bibfield{author}{\bibinfo{person}{Ewart~J De~Visser},
  \bibinfo{person}{Samuel~S Monfort}, \bibinfo{person}{Kimberly Goodyear},
  \bibinfo{person}{Li Lu}, \bibinfo{person}{Martin O'Hara},
  \bibinfo{person}{Mary~R Lee}, \bibinfo{person}{Raja Parasuraman}, {and}
  \bibinfo{person}{Frank Krueger}.} \bibinfo{year}{2017}\natexlab{}.
\newblock \showarticletitle{A Little Anthropomorphism Goes a Long Way:
  {{Effects}} of Oxytocin on Trust, Compliance, and Team Performance with
  Automated Agents}.
\newblock \bibinfo{journal}{\emph{Human factors}} \bibinfo{volume}{59},
  \bibinfo{number}{1} (\bibinfo{year}{2017}), \bibinfo{pages}{116--133}.
\newblock
\urldef\tempurl%
\url{https://doi.org/10.1177/0018720816687205}
\showDOI{\tempurl}


\bibitem[\protect\citeauthoryear{Dijksterhuis, Hurley, and Chater}{Dijksterhuis
  et~al\mbox{.}}{2005}]%
        {dijksterhuis2005we}
\bibfield{author}{\bibinfo{person}{Ap Dijksterhuis}, \bibinfo{person}{Susan
  Hurley}, {and} \bibinfo{person}{Nick Chater}.}
  \bibinfo{year}{2005}\natexlab{}.
\newblock \showarticletitle{Why We Are Social Animals}.
\newblock In \bibinfo{booktitle}{\emph{Perspectives on Imitation: {{From}}
  Neuroscience to Social Science - Volume 2: {{Imitation}}, Human Development,
  and Culture}}. \bibinfo{series}{{{CogNet}}}, Vol.~\bibinfo{volume}{2}.
  \bibinfo{publisher}{{The MIT Press}}.
\newblock
\showISBNx{978-0-262-27595-8}
\urldef\tempurl%
\url{https://doi.org/10.7551/mitpress/5331.003.0012}
\showDOI{\tempurl}
\showeprint{https://direct.mit.edu/book/chapter-pdf/173158/9780262275958\textbackslash\_caj.pdf}


\bibitem[\protect\citeauthoryear{Ding, Veeman, and Adamowicz}{Ding
  et~al\mbox{.}}{2013}]%
        {Ding2013}
\bibfield{author}{\bibinfo{person}{Yulian Ding}, \bibinfo{person}{Michele~M.
  Veeman}, {and} \bibinfo{person}{Wiktor~L. Adamowicz}.}
  \bibinfo{year}{2013}\natexlab{}.
\newblock \showarticletitle{The Influence of Trust on Consumer Behavior: {{An}}
  Application to Recurring Food Risks in {{Canada}}}.
\newblock \bibinfo{journal}{\emph{Journal of Economic Behavior and
  Organization}}  \bibinfo{volume}{92} (\bibinfo{date}{Aug.}
  \bibinfo{year}{2013}), \bibinfo{pages}{214--223}.
\newblock
\urldef\tempurl%
\url{https://doi.org/10.1016/j.jebo.2013.06.009}
\showDOI{\tempurl}


\bibitem[\protect\citeauthoryear{Dybowski, Raczka, Postarnak, Castro, and
  Silva}{Dybowski et~al\mbox{.}}{2022}]%
        {dybowski2022interpersonal}
\bibfield{author}{\bibinfo{person}{Karol Dybowski}, \bibinfo{person}{Barbara
  Raczka}, \bibinfo{person}{Svetlana Postarnak},
  \bibinfo{person}{S{\~a}o~Lu{\'i}s Castro}, {and} \bibinfo{person}{Susana
  Silva}.} \bibinfo{year}{2022}\natexlab{}.
\newblock \showarticletitle{Interpersonal {{Synchronization Protects Against}}
  the {{Antisocial Outcomes}} of {{Frustration}}}.
\newblock \bibinfo{journal}{\emph{Psychological Reports}} (\bibinfo{date}{Jan.}
  \bibinfo{year}{2022}).
\newblock
\showISSN{0033-2941}
\urldef\tempurl%
\url{https://doi.org/10.1177/00332941211054771}
\showDOI{\tempurl}


\bibitem[\protect\citeauthoryear{Fusaroli, {R{\k{a}}czaszek-Leonardi}, and
  Tyl{\'e}n}{Fusaroli et~al\mbox{.}}{2014}]%
        {fusaroli2014dialog}
\bibfield{author}{\bibinfo{person}{Riccardo Fusaroli}, \bibinfo{person}{Joanna
  {R{\k{a}}czaszek-Leonardi}}, {and} \bibinfo{person}{Kristian Tyl{\'e}n}.}
  \bibinfo{year}{2014}\natexlab{}.
\newblock \showarticletitle{Dialog as Interpersonal Synergy}.
\newblock \bibinfo{journal}{\emph{New Ideas in Psychology}}
  \bibinfo{volume}{32} (\bibinfo{year}{2014}), \bibinfo{pages}{147--157}.
\newblock
\urldef\tempurl%
\url{https://doi.org/10.1016/j.newideapsych.2013.03.005}
\showDOI{\tempurl}


\bibitem[\protect\citeauthoryear{Gompei and Umemuro}{Gompei and
  Umemuro}{2018}]%
        {gompei2018factors}
\bibfield{author}{\bibinfo{person}{Takayuki Gompei} {and}
  \bibinfo{person}{Hiroyuki Umemuro}.} \bibinfo{year}{2018}\natexlab{}.
\newblock \showarticletitle{Factors and {{Development}} of {{Cognitive}} and
  {{Affective Trust}} on {{Social Robots}}}. In
  \bibinfo{booktitle}{\emph{Social {{Robotics}}}}
  \emph{(\bibinfo{series}{Lecture {{Notes}} in {{Computer Science}}})},
  \bibfield{editor}{\bibinfo{person}{Shuzhi~Sam Ge}, \bibinfo{person}{John-John
  Cabibihan}, \bibinfo{person}{Miguel~A. Salichs}, \bibinfo{person}{Elizabeth
  Broadbent}, \bibinfo{person}{Hongsheng He}, \bibinfo{person}{Alan~R. Wagner},
  {and} \bibinfo{person}{{\'A}lvaro {Castro-Gonz{\'a}lez}}} (Eds.).
  \bibinfo{publisher}{{Springer International Publishing}},
  \bibinfo{address}{{Cham}}, \bibinfo{pages}{45--54}.
\newblock
\showISBNx{978-3-030-05204-1}
\urldef\tempurl%
\url{https://doi.org/10.1007/978-3-030-05204-1_5}
\showDOI{\tempurl}


\bibitem[\protect\citeauthoryear{Hancock, Billings, Schaefer, Chen, De~Visser,
  and Parasuraman}{Hancock et~al\mbox{.}}{2011}]%
        {Hancock2011}
\bibfield{author}{\bibinfo{person}{Peter~A. Hancock},
  \bibinfo{person}{Deborah~R. Billings}, \bibinfo{person}{Kristin~E. Schaefer},
  \bibinfo{person}{Jessie~Y.C. Chen}, \bibinfo{person}{Ewart~J. De~Visser},
  {and} \bibinfo{person}{Raja Parasuraman}.} \bibinfo{year}{2011}\natexlab{}.
\newblock \showarticletitle{A Meta-Analysis of Factors Affecting Trust in
  Human-Robot Interaction}.
\newblock \bibinfo{journal}{\emph{Human Factors}} \bibinfo{volume}{53},
  \bibinfo{number}{5} (\bibinfo{date}{Oct.} \bibinfo{year}{2011}),
  \bibinfo{pages}{517--527}.
\newblock
\showISSN{00187208}
\urldef\tempurl%
\url{https://doi.org/10.1177/0018720811417254}
\showDOI{\tempurl}


\bibitem[\protect\citeauthoryear{Hancock, Kessler, Kaplan, Brill, and
  Szalma}{Hancock et~al\mbox{.}}{2021}]%
        {hancock2021evolving}
\bibfield{author}{\bibinfo{person}{Peter~A Hancock}, \bibinfo{person}{Theresa~T
  Kessler}, \bibinfo{person}{Alexandra~D Kaplan}, \bibinfo{person}{John~C
  Brill}, {and} \bibinfo{person}{James~L Szalma}.}
  \bibinfo{year}{2021}\natexlab{}.
\newblock \showarticletitle{Evolving Trust in Robots: Specification through
  Sequential and Comparative Meta-Analyses}.
\newblock \bibinfo{journal}{\emph{Human factors}} \bibinfo{volume}{63},
  \bibinfo{number}{7} (\bibinfo{year}{2021}), \bibinfo{pages}{1196--1229}.
\newblock
\urldef\tempurl%
\url{https://doi.org/10.1177/0018720820922080}
\showDOI{\tempurl}


\bibitem[\protect\citeauthoryear{Hashimoto, Yamano, and Usui}{Hashimoto
  et~al\mbox{.}}{2009}]%
        {hashimoto2009effects}
\bibfield{author}{\bibinfo{person}{Minoru Hashimoto}, \bibinfo{person}{Misaki
  Yamano}, {and} \bibinfo{person}{Tatsuya Usui}.}
  \bibinfo{year}{2009}\natexlab{}.
\newblock \showarticletitle{Effects of Emotional Synchronization in Human-Robot
  {{KANSEI}} Communications}. In \bibinfo{booktitle}{\emph{{{RO-MAN}} 2009 -
  {{The}} 18th {{IEEE International Symposium}} on {{Robot}} and {{Human
  Interactive Communication}}}}. \bibinfo{publisher}{{IEEE Comput. Soc}},
  \bibinfo{address}{{Toyama, Japan}}, \bibinfo{pages}{52--57}.
\newblock
\showISSN{1944-9437}
\urldef\tempurl%
\url{https://doi.org/10.1109/ROMAN.2009.5326232}
\showDOI{\tempurl}


\bibitem[\protect\citeauthoryear{Hida}{Hida}{1980}]%
        {hida1980brownian}
\bibfield{author}{\bibinfo{person}{T. Hida}.} \bibinfo{year}{1980}\natexlab{}.
\newblock \showarticletitle{Brownian {{Motion}}}.
\newblock In \bibinfo{booktitle}{\emph{Brownian {{Motion}}}},
  \bibfield{editor}{\bibinfo{person}{T.~Hida}} (Ed.).
  \bibinfo{publisher}{{Springer US}}, \bibinfo{address}{{New York, NY}},
  \bibinfo{pages}{44--113}.
\newblock
\showISBNx{978-1-4612-6030-1}
\urldef\tempurl%
\url{https://doi.org/10.1007/978-1-4612-6030-1_2}
\showDOI{\tempurl}


\bibitem[\protect\citeauthoryear{Hoffman and Weinberg}{Hoffman and
  Weinberg}{2011}]%
        {hoffman2011interactive}
\bibfield{author}{\bibinfo{person}{Guy Hoffman} {and} \bibinfo{person}{Gil
  Weinberg}.} \bibinfo{year}{2011}\natexlab{}.
\newblock \showarticletitle{Interactive Improvisation with a Robotic Marimba
  Player}.
\newblock \bibinfo{journal}{\emph{Autonomous Robots}} \bibinfo{volume}{31},
  \bibinfo{number}{2} (\bibinfo{year}{2011}), \bibinfo{pages}{133--153}.
\newblock
\urldef\tempurl%
\url{https://doi.org/10.1007/s10514-011-9237-0}
\showDOI{\tempurl}


\bibitem[\protect\citeauthoryear{Hofree, Ruvolo, Bartlett, and
  Winkielman}{Hofree et~al\mbox{.}}{2014}]%
        {hofree2014bridging}
\bibfield{author}{\bibinfo{person}{Galit Hofree}, \bibinfo{person}{Paul
  Ruvolo}, \bibinfo{person}{Marian~Stewart Bartlett}, {and}
  \bibinfo{person}{Piotr Winkielman}.} \bibinfo{year}{2014}\natexlab{}.
\newblock \showarticletitle{Bridging the Mechanical and the Human Mind:
  Spontaneous Mimicry of a Physically Present Android}.
\newblock \bibinfo{journal}{\emph{PloS one}} \bibinfo{volume}{9},
  \bibinfo{number}{7} (\bibinfo{year}{2014}), \bibinfo{pages}{e99934}.
\newblock
\urldef\tempurl%
\url{https://doi.org/10.1371/journal.pone.0099934}
\showDOI{\tempurl}


\bibitem[\protect\citeauthoryear{Hove and Risen}{Hove and Risen}{2009}]%
        {hove2009s}
\bibfield{author}{\bibinfo{person}{Michael~J Hove} {and}
  \bibinfo{person}{Jane~L Risen}.} \bibinfo{year}{2009}\natexlab{}.
\newblock \showarticletitle{It's All in the Timing: {{Interpersonal}} Synchrony
  Increases Affiliation}.
\newblock \bibinfo{journal}{\emph{Social cognition}} \bibinfo{volume}{27},
  \bibinfo{number}{6} (\bibinfo{year}{2009}), \bibinfo{pages}{949--960}.
\newblock
\urldef\tempurl%
\url{https://doi.org/10.1521/soco.2009.27.6.949}
\showDOI{\tempurl}


\bibitem[\protect\citeauthoryear{Jian, Bisantz, and Drury}{Jian
  et~al\mbox{.}}{2000}]%
        {Jian2000}
\bibfield{author}{\bibinfo{person}{Jiun-Yin Jian}, \bibinfo{person}{Ann~M.
  Bisantz}, {and} \bibinfo{person}{Colin~G. Drury}.}
  \bibinfo{year}{2000}\natexlab{}.
\newblock \showarticletitle{Foundations for an {{Empirically Determined Scale}}
  of {{Trust}} in {{Automated Systems}}}.
\newblock \bibinfo{journal}{\emph{International Journal of Cognitive
  Ergonomics}} \bibinfo{volume}{4}, \bibinfo{number}{1} (\bibinfo{date}{March}
  \bibinfo{year}{2000}), \bibinfo{pages}{53--71}.
\newblock
\showISSN{1088-6362}
\urldef\tempurl%
\url{https://doi.org/10.1207/S15327566IJCE0401_04}
\showDOI{\tempurl}


\bibitem[\protect\citeauthoryear{Kirkpatrick, Hahn, and Haufler}{Kirkpatrick
  et~al\mbox{.}}{2017}]%
        {kirkpatrick201710}
\bibfield{author}{\bibinfo{person}{Jesse Kirkpatrick}, \bibinfo{person}{Erin~N.
  Hahn}, {and} \bibinfo{person}{Amy~J. Haufler}.}
  \bibinfo{year}{2017}\natexlab{}.
\newblock \showarticletitle{Trust and {{Human}}\textendash{{Robot
  Interactions}}}.
\newblock \bibinfo{journal}{\emph{Robot ethics 2.0: from autonomous cars to
  artificial intelligence}}  \bibinfo{volume}{1} (\bibinfo{date}{Oct.}
  \bibinfo{year}{2017}), \bibinfo{pages}{91}.
\newblock
\urldef\tempurl%
\url{https://doi.org/10.1093/oso/9780190652951.003.0010}
\showDOI{\tempurl}


\bibitem[\protect\citeauthoryear{Kohn, {de Visser}, Wiese, Lee, and Shaw}{Kohn
  et~al\mbox{.}}{2021}]%
        {kohn2021measurement}
\bibfield{author}{\bibinfo{person}{Spencer~C. Kohn}, \bibinfo{person}{Ewart~J.
  {de Visser}}, \bibinfo{person}{Eva Wiese}, \bibinfo{person}{Yi-Ching Lee},
  {and} \bibinfo{person}{Tyler~H. Shaw}.} \bibinfo{year}{2021}\natexlab{}.
\newblock \showarticletitle{Measurement of {{Trust}} in {{Automation}}: {{A
  Narrative Review}} and {{Reference Guide}}}.
\newblock \bibinfo{journal}{\emph{Frontiers in Psychology}}
  \bibinfo{volume}{12} (\bibinfo{year}{2021}), \bibinfo{pages}{604977}.
\newblock
\showISSN{1664-1078}
\urldef\tempurl%
\url{https://doi.org/10.3389/fpsyg.2021.604977}
\showDOI{\tempurl}


\bibitem[\protect\citeauthoryear{Kok and Soh}{Kok and Soh}{2020}]%
        {Kok2020}
\bibfield{author}{\bibinfo{person}{Bing~Cai Kok} {and} \bibinfo{person}{Harold
  Soh}.} \bibinfo{year}{2020}\natexlab{}.
\newblock \showarticletitle{Trust in {{Robots}}: {{Challenges}} and
  {{Opportunities}}}.
\newblock \bibinfo{journal}{\emph{Current Robotics Reports}}
  \bibinfo{volume}{1}, \bibinfo{number}{4} (\bibinfo{date}{Dec.}
  \bibinfo{year}{2020}), \bibinfo{pages}{297--309}.
\newblock
\showISBNx{4315402000029}
\showISSN{2662-4087}
\urldef\tempurl%
\url{https://doi.org/10.1007/s43154-020-00029-y}
\showDOI{\tempurl}


\bibitem[\protect\citeauthoryear{{Kose-Bagci}, Ferrari, Dautenhahn, Syrdal, and
  Nehaniv}{{Kose-Bagci} et~al\mbox{.}}{2009}]%
        {kose2009effects}
\bibfield{author}{\bibinfo{person}{Hatice {Kose-Bagci}}, \bibinfo{person}{Ester
  Ferrari}, \bibinfo{person}{Kerstin Dautenhahn}, \bibinfo{person}{Dag~Sverre
  Syrdal}, {and} \bibinfo{person}{Chrystopher~L Nehaniv}.}
  \bibinfo{year}{2009}\natexlab{}.
\newblock \showarticletitle{Effects of Embodiment and Gestures on Social
  Interaction in Drumming Games with a Humanoid Robot}.
\newblock \bibinfo{journal}{\emph{Advanced Robotics}} \bibinfo{volume}{23},
  \bibinfo{number}{14} (\bibinfo{year}{2009}), \bibinfo{pages}{1951--1996}.
\newblock
\urldef\tempurl%
\url{https://doi.org/10.1163/016918609X12518783330360}
\showDOI{\tempurl}


\bibitem[\protect\citeauthoryear{Laird}{Laird}{1974}]%
        {laird1974self}
\bibfield{author}{\bibinfo{person}{James~D Laird}.}
  \bibinfo{year}{1974}\natexlab{}.
\newblock \showarticletitle{Self-Attribution of Emotion: The Effects of
  Expressive Behavior on the Quality of Emotional Experience.}
\newblock \bibinfo{journal}{\emph{Journal of personality and social
  psychology}} \bibinfo{volume}{29}, \bibinfo{number}{4}
  (\bibinfo{year}{1974}), \bibinfo{pages}{475}.
\newblock
\urldef\tempurl%
\url{https://doi.org/10.1037/h0036125}
\showDOI{\tempurl}


\bibitem[\protect\citeauthoryear{Lakin and Chartrand}{Lakin and
  Chartrand}{2003}]%
        {lakin2003using}
\bibfield{author}{\bibinfo{person}{Jessica~L Lakin} {and}
  \bibinfo{person}{Tanya~L Chartrand}.} \bibinfo{year}{2003}\natexlab{}.
\newblock \showarticletitle{Using Nonconscious Behavioral Mimicry to Create
  Affiliation and Rapport}.
\newblock \bibinfo{journal}{\emph{Psychological science}} \bibinfo{volume}{14},
  \bibinfo{number}{4} (\bibinfo{year}{2003}), \bibinfo{pages}{334--339}.
\newblock
\urldef\tempurl%
\url{https://doi.org/10.1111/1467-9280.14481}
\showDOI{\tempurl}


\bibitem[\protect\citeauthoryear{Langer, {Feingold-Polak}, Mueller, Kellmeyer,
  and {Levy-Tzedek}}{Langer et~al\mbox{.}}{2019}]%
        {Langer2019}
\bibfield{author}{\bibinfo{person}{Allison Langer}, \bibinfo{person}{R.
  {Feingold-Polak}}, \bibinfo{person}{Oliver Mueller}, \bibinfo{person}{Philipp
  Kellmeyer}, {and} \bibinfo{person}{Shelly {Levy-Tzedek}}.}
  \bibinfo{year}{2019}\natexlab{}.
\newblock \showarticletitle{Trust in Socially Assistive Robots:
  {{Considerations}} for Use in Rehabilitation}.
\newblock \bibinfo{journal}{\emph{Neuroscience \textbackslash\& Biobehavioral
  Reviews}}  \bibinfo{volume}{104} (\bibinfo{date}{Sept.}
  \bibinfo{year}{2019}), \bibinfo{pages}{231--239}.
\newblock
\showISSN{0149-7634}
\urldef\tempurl%
\url{https://doi.org/10.1016/J.NEUBIOREV.2019.07.014}
\showDOI{\tempurl}


\bibitem[\protect\citeauthoryear{Launay, Dean, and Bailes}{Launay
  et~al\mbox{.}}{2013}]%
        {launay2013synchronization}
\bibfield{author}{\bibinfo{person}{Jacques Launay}, \bibinfo{person}{Roger~T
  Dean}, {and} \bibinfo{person}{Freya Bailes}.}
  \bibinfo{year}{2013}\natexlab{}.
\newblock \showarticletitle{Synchronization Can Influence Trust Following
  Virtual Interaction.}
\newblock \bibinfo{journal}{\emph{Experimental psychology}}
  \bibinfo{volume}{60}, \bibinfo{number}{1} (\bibinfo{year}{2013}),
  \bibinfo{pages}{53}.
\newblock
\urldef\tempurl%
\url{https://doi.org/10.1027/1618-3169/a000173}
\showDOI{\tempurl}


\bibitem[\protect\citeauthoryear{Lestienne}{Lestienne}{2001}]%
        {Lestienne2001}
\bibfield{author}{\bibinfo{person}{R{\'e}my Lestienne}.}
  \bibinfo{year}{2001}\natexlab{}.
\newblock \showarticletitle{Spike Timing, Synchronization and Information
  Processing on the Sensory Side of the Central Nervous System}.
\newblock \bibinfo{journal}{\emph{Progress in Neurobiology}}
  \bibinfo{volume}{65}, \bibinfo{number}{6} (\bibinfo{date}{Dec.}
  \bibinfo{year}{2001}), \bibinfo{pages}{545--591}.
\newblock
\urldef\tempurl%
\url{https://doi.org/10.1016/s0301-0082(01)00019-3}
\showDOI{\tempurl}


\bibitem[\protect\citeauthoryear{Malhotra and Lumineau}{Malhotra and
  Lumineau}{2011}]%
        {Malhotra2011}
\bibfield{author}{\bibinfo{person}{Deepak Malhotra} {and}
  \bibinfo{person}{Fabrice Lumineau}.} \bibinfo{year}{2011}\natexlab{}.
\newblock \showarticletitle{Trust and Collaboration in the Aftermath of
  Conflict: {{The}} Effects of Contract Structure}.
\newblock \bibinfo{journal}{\emph{Academy of Management Journal}}
  \bibinfo{volume}{54}, \bibinfo{number}{5} (\bibinfo{date}{Oct.}
  \bibinfo{year}{2011}), \bibinfo{pages}{981--998}.
\newblock
\urldef\tempurl%
\url{https://doi.org/10.5465/amj.2009.0683}
\showDOI{\tempurl}


\bibitem[\protect\citeauthoryear{Marsh, Richardson, and Schmidt}{Marsh
  et~al\mbox{.}}{2009}]%
        {marsh2009social}
\bibfield{author}{\bibinfo{person}{Kerry~L Marsh}, \bibinfo{person}{Michael~J
  Richardson}, {and} \bibinfo{person}{Richard~C Schmidt}.}
  \bibinfo{year}{2009}\natexlab{}.
\newblock \showarticletitle{Social Connection through Joint Action and
  Interpersonal Coordination}.
\newblock \bibinfo{journal}{\emph{Topics in Cognitive Science}}
  \bibinfo{volume}{1}, \bibinfo{number}{2} (\bibinfo{year}{2009}),
  \bibinfo{pages}{320--339}.
\newblock
\urldef\tempurl%
\url{https://doi.org/10.1111/j.1756-8765.2009.01022.x}
\showDOI{\tempurl}


\bibitem[\protect\citeauthoryear{Miles, Nind, and Macrae}{Miles
  et~al\mbox{.}}{2009}]%
        {miles2009rhythm}
\bibfield{author}{\bibinfo{person}{Lynden~K Miles}, \bibinfo{person}{Louise~K
  Nind}, {and} \bibinfo{person}{C~Neil Macrae}.}
  \bibinfo{year}{2009}\natexlab{}.
\newblock \showarticletitle{The Rhythm of Rapport: {{Interpersonal}} Synchrony
  and Social Perception}.
\newblock \bibinfo{journal}{\emph{Journal of experimental social psychology}}
  \bibinfo{volume}{45}, \bibinfo{number}{3} (\bibinfo{year}{2009}),
  \bibinfo{pages}{585--589}.
\newblock
\urldef\tempurl%
\url{https://doi.org/10.1016/j.jesp.2009.02.002}
\showDOI{\tempurl}


\bibitem[\protect\citeauthoryear{M{\"o}rtl, Lorenz, and Hirche}{M{\"o}rtl
  et~al\mbox{.}}{2014}]%
        {mortl2014rhythm}
\bibfield{author}{\bibinfo{person}{Alexander M{\"o}rtl},
  \bibinfo{person}{Tamara Lorenz}, {and} \bibinfo{person}{Sandra Hirche}.}
  \bibinfo{year}{2014}\natexlab{}.
\newblock \showarticletitle{Rhythm Patterns Interaction-Synchronization
  Behavior for Human-Robot Joint Action}.
\newblock \bibinfo{journal}{\emph{PloS one}} \bibinfo{volume}{9},
  \bibinfo{number}{4} (\bibinfo{year}{2014}), \bibinfo{pages}{e95195}.
\newblock
\urldef\tempurl%
\url{https://doi.org/10.1371/journal.pone.0095195}
\showDOI{\tempurl}


\bibitem[\protect\citeauthoryear{Mota, Rea, Le~Tran, Young, Sharlin, and
  Sousa}{Mota et~al\mbox{.}}{2016}]%
        {mota2016playing}
\bibfield{author}{\bibinfo{person}{Roberta C.~Ramos Mota},
  \bibinfo{person}{Daniel~J. Rea}, \bibinfo{person}{Anna Le~Tran},
  \bibinfo{person}{James~E. Young}, \bibinfo{person}{Ehud Sharlin}, {and}
  \bibinfo{person}{Mario~C. Sousa}.} \bibinfo{year}{2016}\natexlab{}.
\newblock \showarticletitle{Playing the `Trust Game' with Robots: {{Social}}
  Strategies and Experiences}. In \bibinfo{booktitle}{\emph{2016 25th {{IEEE
  International Symposium}} on {{Robot}} and {{Human Interactive
  Communication}} ({{RO-MAN}})}}. \bibinfo{publisher}{{IEEE Comput. Soc}},
  \bibinfo{address}{{Columbia University, NY, USA}}, \bibinfo{pages}{519--524}.
\newblock
\showISSN{1944-9437}
\urldef\tempurl%
\url{https://doi.org/10.1109/ROMAN.2016.7745167}
\showDOI{\tempurl}


\bibitem[\protect\citeauthoryear{Naneva, Sarda~Gou, Webb, and Prescott}{Naneva
  et~al\mbox{.}}{2020}]%
        {naneva2020systematic}
\bibfield{author}{\bibinfo{person}{Stanislava Naneva}, \bibinfo{person}{Marina
  Sarda~Gou}, \bibinfo{person}{Thomas~L. Webb}, {and} \bibinfo{person}{Tony~J.
  Prescott}.} \bibinfo{year}{2020}\natexlab{}.
\newblock \showarticletitle{A {{Systematic Review}} of {{Attitudes}},
  {{Anxiety}}, {{Acceptance}}, and {{Trust Towards Social Robots}}}.
\newblock \bibinfo{journal}{\emph{International Journal of Social Robotics}}
  \bibinfo{volume}{12}, \bibinfo{number}{6} (\bibinfo{date}{Dec.}
  \bibinfo{year}{2020}), \bibinfo{pages}{1179--1201}.
\newblock
\showISSN{1875-4805}
\urldef\tempurl%
\url{https://doi.org/10.1007/s12369-020-00659-4}
\showDOI{\tempurl}


\bibitem[\protect\citeauthoryear{Nazari, Castro, and Godage}{Nazari
  et~al\mbox{.}}{2019}]%
        {nazari2019forward}
\bibfield{author}{\bibinfo{person}{Ali~A. Nazari}, \bibinfo{person}{Diego
  Castro}, {and} \bibinfo{person}{Isuru~S. Godage}.}
  \bibinfo{year}{2019}\natexlab{}.
\newblock \bibinfo{title}{Forward and {{Inverse Kinematics}} of a {{Single
  Section Inextensible Continuum Arm}}}.
\newblock
\newblock
\urldef\tempurl%
\url{https://doi.org/10.48550/arXiv.1907.06518}
\showDOI{\tempurl}
\showeprint[arxiv]{1907.06518}~[cs]


\bibitem[\protect\citeauthoryear{Newtson}{Newtson}{1994}]%
        {newtson1994perception}
\bibfield{author}{\bibinfo{person}{Darren Newtson}.}
  \bibinfo{year}{1994}\natexlab{}.
\newblock \showarticletitle{The Perception and Coupling of Behavior Waves}.
\newblock In \bibinfo{booktitle}{\emph{Dynamical Systems in Social
  Psychology.}} \bibinfo{publisher}{{Academic Press}}, \bibinfo{address}{{San
  Diego, CA, US}}, \bibinfo{pages}{139--167}.
\newblock
\showISBNx{0-12-709990-5 (Hardcover)}


\bibitem[\protect\citeauthoryear{Nowak, Vallacher, Borkowski,
  et~al\mbox{.}}{Nowak et~al\mbox{.}}{2000}]%
        {nowak2000modeling}
\bibfield{author}{\bibinfo{person}{Andrzej Nowak}, \bibinfo{person}{Robin~R
  Vallacher}, \bibinfo{person}{Wojciech Borkowski}, {et~al\mbox{.}}}
  \bibinfo{year}{2000}\natexlab{}.
\newblock \showarticletitle{Modeling the Temporal Coordination of Behavior and
  Internal States}.
\newblock \bibinfo{journal}{\emph{Advances in Complex Systems}}
  \bibinfo{volume}{3}, \bibinfo{number}{1-4} (\bibinfo{year}{2000}),
  \bibinfo{pages}{67--86}.
\newblock
\urldef\tempurl%
\url{https://doi.org/10.1142/S0219525900000066}
\showDOI{\tempurl}


\bibitem[\protect\citeauthoryear{Nowak, Vallacher, and Burnstein}{Nowak
  et~al\mbox{.}}{1998}]%
        {nowak1998computational}
\bibfield{author}{\bibinfo{person}{Andrzej Nowak}, \bibinfo{person}{Robin~R.
  Vallacher}, {and} \bibinfo{person}{Eugene Burnstein}.}
  \bibinfo{year}{1998}\natexlab{}.
\newblock \showarticletitle{Computational Social Psychology: {{A}} Neural
  Network Approach to Interpersonal Dynamics}.
\newblock In \bibinfo{booktitle}{\emph{Computer Modeling of Social Processes}}.
  \bibinfo{publisher}{{Sage Publications Ltd}}, \bibinfo{address}{{Thousand
  Oaks, CA}}, \bibinfo{pages}{97--125}.
\newblock
\showISBNx{978-0-7619-5423-1 978-0-7619-5424-8}


\bibitem[\protect\citeauthoryear{Nowak, Vallacher, Zochowski, and
  Rychwalska}{Nowak et~al\mbox{.}}{2017}]%
        {Nowak2017}
\bibfield{author}{\bibinfo{person}{Andrzej Nowak}, \bibinfo{person}{Robin~R.
  Vallacher}, \bibinfo{person}{Michal Zochowski}, {and}
  \bibinfo{person}{Agnieszka Rychwalska}.} \bibinfo{year}{2017}\natexlab{}.
\newblock \showarticletitle{Functional Synchronization: {{The}} Emergence of
  Coordinated Activity in Human Systems}.
\newblock \bibinfo{journal}{\emph{Frontiers in Psychology}}
  \bibinfo{volume}{8}, \bibinfo{number}{JUN} (\bibinfo{date}{June}
  \bibinfo{year}{2017}), \bibinfo{pages}{945}.
\newblock
\showISSN{16641078}
\urldef\tempurl%
\url{https://doi.org/10.3389/FPSYG.2017.00945}
\showDOI{\tempurl}


\bibitem[\protect\citeauthoryear{Nowak, Vallacher, Praszkier, Rychwalska, and
  Zochowski}{Nowak et~al\mbox{.}}{2020}]%
        {Nowak2020}
\bibfield{author}{\bibinfo{person}{Andrzej~K Nowak}, \bibinfo{person}{Robin~R
  Vallacher}, \bibinfo{person}{Ryszard Praszkier}, \bibinfo{person}{Agnieszka
  Rychwalska}, {and} \bibinfo{person}{Michal Zochowski}.}
  \bibinfo{year}{2020}\natexlab{}.
\newblock \bibinfo{booktitle}{\emph{In Sync: {{The}} Emergence of Function in
  Minds, Groups and Societies}}.
\newblock \bibinfo{publisher}{{Springer Cham}},
  \bibinfo{address}{{Switzerland}}.
\newblock
\showISBNx{978-3-030-38987-1}


\bibitem[\protect\citeauthoryear{Oksanen, Savela, Latikka, and Koivula}{Oksanen
  et~al\mbox{.}}{2020}]%
        {oksanen2020trust}
\bibfield{author}{\bibinfo{person}{Atte Oksanen}, \bibinfo{person}{Nina
  Savela}, \bibinfo{person}{Rita Latikka}, {and} \bibinfo{person}{Aki
  Koivula}.} \bibinfo{year}{2020}\natexlab{}.
\newblock \showarticletitle{Trust toward Robots and Artificial Intelligence:
  {{An}} Experimental Approach to Human\textendash Technology Interactions
  Online}.
\newblock \bibinfo{journal}{\emph{Frontiers in Psychology}}
  \bibinfo{volume}{11} (\bibinfo{year}{2020}), \bibinfo{pages}{568256}.
\newblock
\urldef\tempurl%
\url{https://doi.org/10.3389/fpsyg.2020.568256}
\showDOI{\tempurl}


\bibitem[\protect\citeauthoryear{Olaronke, Oluwaseun, and Rhoda}{Olaronke
  et~al\mbox{.}}{2017}]%
        {Olaronke2017}
\bibfield{author}{\bibinfo{person}{Iroju Olaronke}, \bibinfo{person}{Ojerinde
  Oluwaseun}, {and} \bibinfo{person}{Ikono Rhoda}.}
  \bibinfo{year}{2017}\natexlab{}.
\newblock \showarticletitle{State {{Of The Art}}: {{A Study}} of {{Human-Robot
  Interaction}} in {{Healthcare}}}.
\newblock \bibinfo{journal}{\emph{Information Engineering and Electronic
  Business}}  \bibinfo{volume}{3} (\bibinfo{year}{2017}),
  \bibinfo{pages}{43--55}.
\newblock
\urldef\tempurl%
\url{https://doi.org/10.5815/ijieeb.2017.03.06}
\showDOI{\tempurl}


\bibitem[\protect\citeauthoryear{Pasquali, {Gonzalez-Billandon}, Rea, Sandini,
  and Sciutti}{Pasquali et~al\mbox{.}}{2021}]%
        {Pasquali2021}
\bibfield{author}{\bibinfo{person}{Dario Pasquali}, \bibinfo{person}{Jonas
  {Gonzalez-Billandon}}, \bibinfo{person}{Francesco Rea},
  \bibinfo{person}{Giulio Sandini}, {and} \bibinfo{person}{Alessandra
  Sciutti}.} \bibinfo{year}{2021}\natexlab{}.
\newblock \showarticletitle{Magic {{iCub}}: {{A Humanoid Robot Autonomously
  Catching}} Your {{Lies}} in a {{Card Game}}}. In
  \bibinfo{booktitle}{\emph{Proceedings of the 2021 {{ACM}}/{{IEEE
  International Conference}} on {{Human-Robot Interaction}}}}
  \emph{(\bibinfo{series}{{{HRI}} '21})}. \bibinfo{publisher}{{Association for
  Computing Machinery}}, \bibinfo{address}{{New York, NY, USA}},
  \bibinfo{pages}{293--302}.
\newblock
\showISBNx{978-1-4503-8289-2}
\urldef\tempurl%
\url{https://doi.org/10.1145/3434073.3444682}
\showDOI{\tempurl}


\bibitem[\protect\citeauthoryear{Pikovsky, Rosenblum, and Kurths}{Pikovsky
  et~al\mbox{.}}{2001}]%
        {pikovsky2001synchronization}
\bibfield{author}{\bibinfo{person}{Arkady Pikovsky}, \bibinfo{person}{Michael
  Rosenblum}, {and} \bibinfo{person}{J{\"u}rgen Kurths}.}
  \bibinfo{year}{2001}\natexlab{}.
\newblock \bibinfo{booktitle}{\emph{Synchronization: {{A}} Universal Concept}}.
  \bibinfo{series}{Cambridge Nonlinear Science Series},
  Vol.~\bibinfo{volume}{2}.
\newblock \bibinfo{publisher}{{Cambridge University Press}},
  \bibinfo{address}{{Cambridge, England}}.
\newblock


\bibitem[\protect\citeauthoryear{Plaks, Rodriguez, and Ayad}{Plaks
  et~al\mbox{.}}{2022}]%
        {plaks2022identifying}
\bibfield{author}{\bibinfo{person}{Jason~E Plaks},
  \bibinfo{person}{Laura~Bustos Rodriguez}, {and} \bibinfo{person}{Reem Ayad}.}
  \bibinfo{year}{2022}\natexlab{}.
\newblock \showarticletitle{Identifying Psychological Features of Robots That
  Encourage and Discourage Trust}.
\newblock \bibinfo{journal}{\emph{Computers in Human Behavior}}
  \bibinfo{volume}{134} (\bibinfo{year}{2022}), \bibinfo{pages}{107301}.
\newblock
\urldef\tempurl%
\url{https://doi.org/10.1016/j.chb.2022.107301}
\showDOI{\tempurl}


\bibitem[\protect\citeauthoryear{PytlikZillig and Kimbrough}{PytlikZillig and
  Kimbrough}{2016}]%
        {pytlikzillig2016consensus}
\bibfield{author}{\bibinfo{person}{Lisa~M. PytlikZillig} {and}
  \bibinfo{person}{Christopher~D. Kimbrough}.} \bibinfo{year}{2016}\natexlab{}.
\newblock \showarticletitle{Consensus on {{Conceptualizations}} and
  {{Definitions}} of {{Trust}}: {{Are We There Yet}}?}
\newblock In \bibinfo{booktitle}{\emph{Interdisciplinary {{Perspectives}} on
  {{Trust}}: {{Towards Theoretical}} and {{Methodological Integration}}}},
  \bibfield{editor}{\bibinfo{person}{Ellie Shockley},
  \bibinfo{person}{Tess~M.S. Neal}, \bibinfo{person}{Lisa~M. PytlikZillig},
  {and} \bibinfo{person}{Brian~H. Bornstein}} (Eds.).
  \bibinfo{publisher}{{Springer International Publishing}},
  \bibinfo{address}{{Cham}}, \bibinfo{pages}{17--47}.
\newblock
\showISBNx{978-3-319-22261-5}
\urldef\tempurl%
\url{https://doi.org/10.1007/978-3-319-22261-5_2}
\showDOI{\tempurl}


\bibitem[\protect\citeauthoryear{Ravreby, Shilat, and Yeshurun}{Ravreby
  et~al\mbox{.}}{2022}]%
        {ravreby2022liking}
\bibfield{author}{\bibinfo{person}{Inbal Ravreby}, \bibinfo{person}{Yoel
  Shilat}, {and} \bibinfo{person}{Yaara Yeshurun}.}
  \bibinfo{year}{2022}\natexlab{}.
\newblock \showarticletitle{Liking as a Balance between Synchronization,
  Complexity and Novelty}.
\newblock \bibinfo{journal}{\emph{Scientific Reports}} \bibinfo{volume}{12},
  \bibinfo{number}{1} (\bibinfo{year}{2022}), \bibinfo{pages}{1--12}.
\newblock
\urldef\tempurl%
\url{https://doi.org/10.1038/s41598-022-06610-z}
\showDOI{\tempurl}


\bibitem[\protect\citeauthoryear{Richardson}{Richardson}{2016}]%
        {Richardson2016}
\bibfield{author}{\bibinfo{person}{Kathleen Richardson}.}
  \bibinfo{year}{2016}\natexlab{}.
\newblock \showarticletitle{Sex {{Robot Matters}}: {{Slavery}}, the
  {{Prostituted}}, and the {{Rights}} of {{Machines}}}.
\newblock \bibinfo{journal}{\emph{IEEE Technology and Society Magazine}}
  \bibinfo{volume}{35}, \bibinfo{number}{2} (\bibinfo{date}{June}
  \bibinfo{year}{2016}), \bibinfo{pages}{46--53}.
\newblock
\showISSN{0278-0097}
\urldef\tempurl%
\url{https://doi.org/10.1109/MTS.2016.2554421}
\showDOI{\tempurl}


\bibitem[\protect\citeauthoryear{Riegelsberger, Sasse, and
  McCarthy}{Riegelsberger et~al\mbox{.}}{2003}]%
        {riegelsberger2003researcher}
\bibfield{author}{\bibinfo{person}{Jens Riegelsberger},
  \bibinfo{person}{M~Angela Sasse}, {and} \bibinfo{person}{John~D McCarthy}.}
  \bibinfo{year}{2003}\natexlab{}.
\newblock \showarticletitle{The Researcher's Dilemma: Evaluating Trust in
  Computer-Mediated Communication}.
\newblock \bibinfo{journal}{\emph{International Journal of Human-Computer
  Studies}} \bibinfo{volume}{58}, \bibinfo{number}{6} (\bibinfo{year}{2003}),
  \bibinfo{pages}{759--781}.
\newblock
\urldef\tempurl%
\url{https://doi.org/10.1016/S1071-5819(03)00042-9}
\showDOI{\tempurl}


\bibitem[\protect\citeauthoryear{Rinott and Tractinsky}{Rinott and
  Tractinsky}{2022}]%
        {rinott2022designing}
\bibfield{author}{\bibinfo{person}{Michal Rinott} {and} \bibinfo{person}{Noam
  Tractinsky}.} \bibinfo{year}{2022}\natexlab{}.
\newblock \showarticletitle{Designing for Interpersonal Motor Synchronization}.
\newblock \bibinfo{journal}{\emph{Human\textendash Computer Interaction}}
  \bibinfo{volume}{37}, \bibinfo{number}{1} (\bibinfo{year}{2022}),
  \bibinfo{pages}{69--116}.
\newblock
\urldef\tempurl%
\url{https://doi.org/10.1080/07370024.2021.1912608}
\showDOI{\tempurl}


\bibitem[\protect\citeauthoryear{Robinson and Davies}{Robinson and
  Davies}{1999}]%
        {Robinson1999ContinuumR}
\bibfield{author}{\bibinfo{person}{Graham Robinson} {and}
  \bibinfo{person}{J.~Bruce~C. Davies}.} \bibinfo{year}{1999}\natexlab{}.
\newblock \showarticletitle{Continuum Robots - a State of the Art}.
\newblock \bibinfo{journal}{\emph{Proceedings 1999 IEEE International
  Conference on Robotics and Automation (Cat. No.99CH36288C)}}
  \bibinfo{volume}{4} (\bibinfo{year}{1999}), \bibinfo{pages}{2849--2854
  vol.4}.
\newblock
\urldef\tempurl%
\url{https://doi.org/10.1109/ROBOT.1999.774029}
\showDOI{\tempurl}


\bibitem[\protect\citeauthoryear{Rosenblum, Kurths, Pikovsky, Schafer, Tass,
  and Abel}{Rosenblum et~al\mbox{.}}{1998}]%
        {Rosenblum1998}
\bibfield{author}{\bibinfo{person}{M.G. Rosenblum}, \bibinfo{person}{J.
  Kurths}, \bibinfo{person}{A. Pikovsky}, \bibinfo{person}{C. Schafer},
  \bibinfo{person}{P. Tass}, {and} \bibinfo{person}{H.-H. Abel}.}
  \bibinfo{year}{1998}\natexlab{}.
\newblock \showarticletitle{Synchronization in Noisy Systems and
  Cardiorespiratory Interaction}.
\newblock \bibinfo{journal}{\emph{IEEE Engineering in Medicine and Biology
  Magazine}} \bibinfo{volume}{17}, \bibinfo{number}{6} (\bibinfo{year}{1998}),
  \bibinfo{pages}{46--53}.
\newblock
\urldef\tempurl%
\url{https://doi.org/10.1109/51.731320}
\showDOI{\tempurl}


\bibitem[\protect\citeauthoryear{Rosenthal, Gurney, and Moore}{Rosenthal
  et~al\mbox{.}}{1981}]%
        {Rosenthal1981}
\bibfield{author}{\bibinfo{person}{Doreen~A. Rosenthal},
  \bibinfo{person}{Ross~M. Gurney}, {and} \bibinfo{person}{Susan~M. Moore}.}
  \bibinfo{year}{1981}\natexlab{}.
\newblock \showarticletitle{From Trust on Intimacy: {{A}} New Inventory for
  Examining Eriksons Stages of Psychosocial Development}.
\newblock \bibinfo{journal}{\emph{Journal of Youth and Adolescence}}
  \bibinfo{volume}{10}, \bibinfo{number}{6} (\bibinfo{date}{Dec.}
  \bibinfo{year}{1981}), \bibinfo{pages}{525--537}.
\newblock
\urldef\tempurl%
\url{https://doi.org/10.1007/bf02087944}
\showDOI{\tempurl}


\bibitem[\protect\citeauthoryear{Salem, Lakatos, Amirabdollahian, and
  Dautenhahn}{Salem et~al\mbox{.}}{2015a}]%
        {Salem2015}
\bibfield{author}{\bibinfo{person}{Maha Salem}, \bibinfo{person}{Gabriella
  Lakatos}, \bibinfo{person}{Farshid Amirabdollahian}, {and}
  \bibinfo{person}{Kerstin Dautenhahn}.} \bibinfo{year}{2015}\natexlab{a}.
\newblock \showarticletitle{Towards {{Safe}} and {{Trustworthy Social Robots}}:
  {{Ethical Challenges}} and {{Practical Issues}}}. In
  \bibinfo{booktitle}{\emph{Social {{Robotics}}}}
  \emph{(\bibinfo{series}{Lecture {{Notes}} in {{Computer Science}}})},
  \bibfield{editor}{\bibinfo{person}{Adriana Tapus}, \bibinfo{person}{Elisabeth
  Andr{\'e}}, \bibinfo{person}{Jean-Claude Martin}, \bibinfo{person}{Fran{\c
  c}ois Ferland}, {and} \bibinfo{person}{Mehdi Ammi}} (Eds.).
  \bibinfo{publisher}{{Springer International Publishing}},
  \bibinfo{address}{{Cham}}, \bibinfo{pages}{584--593}.
\newblock
\showISBNx{978-3-319-25554-5}
\urldef\tempurl%
\url{https://doi.org/10.1007/978-3-319-25554-5_58}
\showDOI{\tempurl}


\bibitem[\protect\citeauthoryear{Salem, Lakatos, Amirabdollahian, and
  Dautenhahn}{Salem et~al\mbox{.}}{2015b}]%
        {salem2015would}
\bibfield{author}{\bibinfo{person}{Maha Salem}, \bibinfo{person}{Gabriella
  Lakatos}, \bibinfo{person}{Farshid Amirabdollahian}, {and}
  \bibinfo{person}{Kerstin Dautenhahn}.} \bibinfo{year}{2015}\natexlab{b}.
\newblock \showarticletitle{Would {{You Trust}} a ({{Faulty}}) {{Robot}}?
  {{Effects}} of {{Error}}, {{Task Type}} and {{Personality}} on {{Human-Robot
  Cooperation}} and {{Trust}}}. In \bibinfo{booktitle}{\emph{Proceedings of the
  {{Tenth Annual ACM}}/{{IEEE International Conference}} on {{Human-Robot
  Interaction}}}} \emph{(\bibinfo{series}{{{HRI}} '15})}.
  \bibinfo{publisher}{{Association for Computing Machinery}},
  \bibinfo{address}{{New York, NY, USA}}, \bibinfo{pages}{141--148}.
\newblock
\showISBNx{978-1-4503-2883-8}
\urldef\tempurl%
\url{https://doi.org/10.1145/2696454.2696497}
\showDOI{\tempurl}


\bibitem[\protect\citeauthoryear{Sanders, MacArthur, Volante, Hancock,
  MacGillivray, Shugars, and Hancock}{Sanders et~al\mbox{.}}{2017}]%
        {sanders2017trust}
\bibfield{author}{\bibinfo{person}{Tracy~L. Sanders}, \bibinfo{person}{Keith
  MacArthur}, \bibinfo{person}{William Volante}, \bibinfo{person}{Gabriella
  Hancock}, \bibinfo{person}{Thomas MacGillivray}, \bibinfo{person}{William
  Shugars}, {and} \bibinfo{person}{P.~A. Hancock}.}
  \bibinfo{year}{2017}\natexlab{}.
\newblock \showarticletitle{Trust and {{Prior Experience}} in {{Human-Robot
  Interaction}}}.
\newblock \bibinfo{journal}{\emph{Proceedings of the Human Factors and
  Ergonomics Society Annual Meeting}} \bibinfo{volume}{61}, \bibinfo{number}{1}
  (\bibinfo{date}{Sept.} \bibinfo{year}{2017}), \bibinfo{pages}{1809--1813}.
\newblock
\showISSN{2169-5067}
\urldef\tempurl%
\url{https://doi.org/10.1177/1541931213601934}
\showDOI{\tempurl}


\bibitem[\protect\citeauthoryear{Savery, Rose, and Weinberg}{Savery
  et~al\mbox{.}}{2019}]%
        {savery2019establishing}
\bibfield{author}{\bibinfo{person}{Richard Savery}, \bibinfo{person}{Ryan
  Rose}, {and} \bibinfo{person}{Gil Weinberg}.}
  \bibinfo{year}{2019}\natexlab{}.
\newblock \showarticletitle{Establishing {{Human-Robot Trust}} through
  {{Music-Driven Robotic Emotion Prosody}} and {{Gesture}}}. In
  \bibinfo{booktitle}{\emph{2019 28th {{IEEE International Conference}} on
  {{Robot}} and {{Human Interactive Communication}} ({{RO-MAN}})}}.
  \bibinfo{publisher}{{IEEE Comput. Soc}}, \bibinfo{address}{{New Delhi,
  India}}, \bibinfo{pages}{1--7}.
\newblock
\showISSN{1944-9437}
\urldef\tempurl%
\url{https://doi.org/10.1109/RO-MAN46459.2019.8956386}
\showDOI{\tempurl}


\bibitem[\protect\citeauthoryear{Schmidt and Richardson}{Schmidt and
  Richardson}{2008}]%
        {schmidt2008dynamics}
\bibfield{author}{\bibinfo{person}{Richard~C. Schmidt} {and}
  \bibinfo{person}{Michael~J. Richardson}.} \bibinfo{year}{2008}\natexlab{}.
\newblock \showarticletitle{Dynamics of Interpersonal Coordination}.
\newblock In \bibinfo{booktitle}{\emph{Coordination: {{Neural}}, Behavioral and
  Social Dynamics}}, \bibfield{editor}{\bibinfo{person}{Armin Fuchs} {and}
  \bibinfo{person}{Viktor~K. Jirsa}} (Eds.). \bibinfo{publisher}{{Springer
  Berlin Heidelberg}}, \bibinfo{address}{{Berlin, Heidelberg}},
  \bibinfo{pages}{281--308}.
\newblock
\showISBNx{978-3-540-74479-5}
\urldef\tempurl%
\url{https://doi.org/10.1007/978-3-540-74479-5_14}
\showDOI{\tempurl}


\bibitem[\protect\citeauthoryear{Scissors, Gill, and Gergle}{Scissors
  et~al\mbox{.}}{2008}]%
        {scissors2008linguistic}
\bibfield{author}{\bibinfo{person}{Lauren~E. Scissors},
  \bibinfo{person}{Alastair~J. Gill}, {and} \bibinfo{person}{Darren Gergle}.}
  \bibinfo{year}{2008}\natexlab{}.
\newblock \showarticletitle{Linguistic Mimicry and Trust in Text-Based
  {{CMC}}}. In \bibinfo{booktitle}{\emph{Proceedings of the 2008 {{ACM}}
  Conference on {{Computer}} Supported Cooperative Work}}
  \emph{(\bibinfo{series}{{{CSCW}} '08})}. \bibinfo{publisher}{{Association for
  Computing Machinery}}, \bibinfo{address}{{New York, NY, USA}},
  \bibinfo{pages}{277--280}.
\newblock
\showISBNx{978-1-60558-007-4}
\urldef\tempurl%
\url{https://doi.org/10.1145/1460563.1460608}
\showDOI{\tempurl}


\bibitem[\protect\citeauthoryear{Semin and Cacioppo}{Semin and
  Cacioppo}{2008}]%
        {Semin2008}
\bibfield{author}{\bibinfo{person}{G{\"u}n~R. Semin} {and}
  \bibinfo{person}{John~T. Cacioppo}.} \bibinfo{year}{2008}\natexlab{}.
\newblock \showarticletitle{Grounding {{Social Cognition}}:
  {{Synchronization}}, {{Coordination}}, and {{Co-Regulation}}}.
\newblock In \bibinfo{booktitle}{\emph{Embodied {{Grounding}}: {{Social}},
  {{Cognitive}}, {{Affective}}, and {{Neuroscientific Approaches}}}},
  \bibfield{editor}{\bibinfo{person}{Eliot~R. Smith} {and}
  \bibinfo{person}{G{\"u}n~R. Semin}} (Eds.). \bibinfo{publisher}{{Cambridge
  University Press}}, \bibinfo{address}{{Cambridge}},
  \bibinfo{pages}{119--147}.
\newblock
\showISBNx{978-0-521-88019-0}
\urldef\tempurl%
\url{https://doi.org/10.1017/CBO9780511805837.006}
\showDOI{\tempurl}


\bibitem[\protect\citeauthoryear{Shen, Dautenhahn, Saunders, and Kose}{Shen
  et~al\mbox{.}}{2015}]%
        {shen2015can}
\bibfield{author}{\bibinfo{person}{Qiming Shen}, \bibinfo{person}{Kerstin
  Dautenhahn}, \bibinfo{person}{Joe Saunders}, {and} \bibinfo{person}{Hatice
  Kose}.} \bibinfo{year}{2015}\natexlab{}.
\newblock \showarticletitle{Can Real-Time, Adaptive Human\textendash Robot
  Motor Coordination Improve Humans' Overall Perception of a Robot?}
\newblock \bibinfo{journal}{\emph{IEEE Transactions on Autonomous Mental
  Development}} \bibinfo{volume}{7}, \bibinfo{number}{1}
  (\bibinfo{year}{2015}), \bibinfo{pages}{52--64}.
\newblock
\urldef\tempurl%
\url{https://doi.org/10.1109/TAMD.2015.2398451}
\showDOI{\tempurl}


\bibitem[\protect\citeauthoryear{Shneiderman}{Shneiderman}{2020}]%
        {shneiderman2020bridging}
\bibfield{author}{\bibinfo{person}{Ben Shneiderman}.}
  \bibinfo{year}{2020}\natexlab{}.
\newblock \showarticletitle{Bridging the Gap between Ethics and Practice:
  Guidelines for Reliable, Safe, and Trustworthy Human-Centered {{AI}}
  Systems}.
\newblock \bibinfo{journal}{\emph{ACM Transactions on Interactive Intelligent
  Systems (TiiS)}} \bibinfo{volume}{10}, \bibinfo{number}{4}
  (\bibinfo{year}{2020}), \bibinfo{pages}{1--31}.
\newblock
\urldef\tempurl%
\url{https://doi.org/10.1145/3419764}
\showDOI{\tempurl}


\bibitem[\protect\citeauthoryear{Slov{\'a}k, Nov{\'a}k, Troubil, Holub, and
  Hofer}{Slov{\'a}k et~al\mbox{.}}{2011}]%
        {slovak2011exploring}
\bibfield{author}{\bibinfo{person}{Petr Slov{\'a}k}, \bibinfo{person}{Peter
  Nov{\'a}k}, \bibinfo{person}{Pavel Troubil}, \bibinfo{person}{Petr Holub},
  {and} \bibinfo{person}{Erik~C. Hofer}.} \bibinfo{year}{2011}\natexlab{}.
\newblock \showarticletitle{Exploring Trust in Group-to-Group
  Video-Conferencing}. In \bibinfo{booktitle}{\emph{{{CHI}} '11 {{Extended
  Abstracts}} on {{Human Factors}} in {{Computing Systems}}}}
  \emph{(\bibinfo{series}{{{CHI EA}} '11})}. \bibinfo{publisher}{{Association
  for Computing Machinery}}, \bibinfo{address}{{New York, NY, USA}},
  \bibinfo{pages}{1459--1464}.
\newblock
\showISBNx{978-1-4503-0268-5}
\urldef\tempurl%
\url{https://doi.org/10.1145/1979742.1979791}
\showDOI{\tempurl}


\bibitem[\protect\citeauthoryear{Slov{\'a}k, Tennent, Reeves, and
  Fitzpatrick}{Slov{\'a}k et~al\mbox{.}}{2014}]%
        {slovak2014exploring}
\bibfield{author}{\bibinfo{person}{Petr Slov{\'a}k}, \bibinfo{person}{Paul
  Tennent}, \bibinfo{person}{Stuart Reeves}, {and} \bibinfo{person}{Geraldine
  Fitzpatrick}.} \bibinfo{year}{2014}\natexlab{}.
\newblock \showarticletitle{Exploring Skin Conductance Synchronisation in
  Everyday Interactions}. In \bibinfo{booktitle}{\emph{Proceedings of the 8th
  {{Nordic Conference}} on {{Human-Computer Interaction}}: {{Fun}}, {{Fast}},
  {{Foundational}}}} \emph{(\bibinfo{series}{{{NordiCHI}} '14})}.
  \bibinfo{publisher}{{Association for Computing Machinery}},
  \bibinfo{address}{{New York, NY, USA}}, \bibinfo{pages}{511--520}.
\newblock
\showISBNx{978-1-4503-2542-4}
\urldef\tempurl%
\url{https://doi.org/10.1145/2639189.2639206}
\showDOI{\tempurl}


\bibitem[\protect\citeauthoryear{Stel and Vonk}{Stel and Vonk}{2010}]%
        {stel2010mimicry}
\bibfield{author}{\bibinfo{person}{Mari{\"e}lle Stel} {and}
  \bibinfo{person}{Roos Vonk}.} \bibinfo{year}{2010}\natexlab{}.
\newblock \showarticletitle{Mimicry in Social Interaction: {{Benefits}} for
  Mimickers, Mimickees, and Their Interaction}.
\newblock \bibinfo{journal}{\emph{British journal of psychology}}
  \bibinfo{volume}{101}, \bibinfo{number}{2} (\bibinfo{year}{2010}),
  \bibinfo{pages}{311--323}.
\newblock
\urldef\tempurl%
\url{https://doi.org/10.1348/000712609X465424}
\showDOI{\tempurl}


\bibitem[\protect\citeauthoryear{Strack, Martin, and Stepper}{Strack
  et~al\mbox{.}}{1988}]%
        {strack1988inhibiting}
\bibfield{author}{\bibinfo{person}{Fritz Strack}, \bibinfo{person}{Leonard~L
  Martin}, {and} \bibinfo{person}{Sabine Stepper}.}
  \bibinfo{year}{1988}\natexlab{}.
\newblock \showarticletitle{Inhibiting and Facilitating Conditions of the Human
  Smile: A Nonobtrusive Test of the Facial Feedback Hypothesis.}
\newblock \bibinfo{journal}{\emph{Journal of personality and social
  psychology}} \bibinfo{volume}{54}, \bibinfo{number}{5}
  (\bibinfo{year}{1988}), \bibinfo{pages}{768}.
\newblock
\urldef\tempurl%
\url{https://doi.org/10.1037/0022-3514.54.5.768}
\showDOI{\tempurl}


\bibitem[\protect\citeauthoryear{Su, Lazar, Bardzell, and Bardzell}{Su
  et~al\mbox{.}}{2019}]%
        {Su2019}
\bibfield{author}{\bibinfo{person}{Norman~Makoto Su}, \bibinfo{person}{Amanda
  Lazar}, \bibinfo{person}{Jeffrey Bardzell}, {and} \bibinfo{person}{Shaowen
  Bardzell}.} \bibinfo{year}{2019}\natexlab{}.
\newblock \showarticletitle{Of {{Dolls}} and {{Men}}: {{Anticipating Sexual
  Intimacy}} with {{Robots}}}.
\newblock \bibinfo{journal}{\emph{ACM Transactions on Computer-Human
  Interaction}} \bibinfo{volume}{26}, \bibinfo{number}{3} (\bibinfo{date}{June}
  \bibinfo{year}{2019}), \bibinfo{pages}{1--35}.
\newblock
\showISSN{1073-0516}
\urldef\tempurl%
\url{https://doi.org/10.1145/3301422}
\showDOI{\tempurl}


\bibitem[\protect\citeauthoryear{{Thomas B. Singh}}{{Thomas B. Singh}}{2012}]%
        {doi:10.1080/21515581.2012.708496}
\bibfield{author}{\bibinfo{person}{{Thomas B. Singh}}.}
  \bibinfo{year}{2012}\natexlab{}.
\newblock \showarticletitle{A Social Interactions Perspective on Trust and Its
  Determinants}.
\newblock \bibinfo{journal}{\emph{Journal of Trust Research}}
  \bibinfo{volume}{2}, \bibinfo{number}{2} (\bibinfo{year}{2012}),
  \bibinfo{pages}{107--135}.
\newblock
\urldef\tempurl%
\url{https://doi.org/10.1080/21515581.2012.708496}
\showDOI{\tempurl}
\showeprint{https://doi.org/10.1080/21515581.2012.708496}


\bibitem[\protect\citeauthoryear{Troiano, Wood, and Harteveld}{Troiano
  et~al\mbox{.}}{2020}]%
        {Troiano2020}
\bibfield{author}{\bibinfo{person}{Giovanni~Maria Troiano},
  \bibinfo{person}{Matthew Wood}, {and} \bibinfo{person}{Casper Harteveld}.}
  \bibinfo{year}{2020}\natexlab{}.
\newblock \showarticletitle{"{{And This}}, {{Kids}}, {{Is How I Met Your
  Mother}}": {{Consumerist}}, {{Mundane}}, and {{Uncanny Futures}} with {{Sex
  Robots}}}. In \bibinfo{booktitle}{\emph{Proceedings of the 2020 {{CHI
  Conference}} on {{Human Factors}} in {{Computing Systems}}}}
  \emph{(\bibinfo{series}{{{CHI}} '20})}. \bibinfo{publisher}{{Association for
  Computing Machinery}}, \bibinfo{address}{{New York, NY, USA}},
  \bibinfo{pages}{1--17}.
\newblock
\showISBNx{978-1-4503-6708-0}
\urldef\tempurl%
\url{https://doi.org/10.1145/3313831.3376598}
\showDOI{\tempurl}


\bibitem[\protect\citeauthoryear{Tuisku, Pekkarinen, Hennala, and
  Melkas}{Tuisku et~al\mbox{.}}{2019}]%
        {Tuisku2019}
\bibfield{author}{\bibinfo{person}{Outi Tuisku}, \bibinfo{person}{Satu
  Pekkarinen}, \bibinfo{person}{Lea Hennala}, {and} \bibinfo{person}{Helin{\"a}
  Melkas}.} \bibinfo{year}{2019}\natexlab{}.
\newblock \showarticletitle{``{{Robots}} Do Not Replace a Nurse with a Beating
  Heart'': {{The}} Publicity around a Robotic Innovation in Elderly Care}.
\newblock \bibinfo{journal}{\emph{Information Technology and People}}
  \bibinfo{volume}{32}, \bibinfo{number}{1} (\bibinfo{date}{Jan.}
  \bibinfo{year}{2019}), \bibinfo{pages}{47--67}.
\newblock
\showISSN{09593845}
\urldef\tempurl%
\url{https://doi.org/10.1108/ITP-06-2018-0277}
\showDOI{\tempurl}


\bibitem[\protect\citeauthoryear{Vallacher, Nowak, and Zochowski}{Vallacher
  et~al\mbox{.}}{2005}]%
        {vallacher2005dynamics}
\bibfield{author}{\bibinfo{person}{Robin~R Vallacher}, \bibinfo{person}{Andrzej
  Nowak}, {and} \bibinfo{person}{Michal Zochowski}.}
  \bibinfo{year}{2005}\natexlab{}.
\newblock \showarticletitle{Dynamics of Social Coordination: {{The}}
  Synchronization of Internal States in Close Relationships}.
\newblock \bibinfo{journal}{\emph{Interaction Studies}} \bibinfo{volume}{6},
  \bibinfo{number}{1} (\bibinfo{year}{2005}), \bibinfo{pages}{35--52}.
\newblock
\urldef\tempurl%
\url{https://doi.org/10.1075/is.6.1.04val}
\showDOI{\tempurl}


\bibitem[\protect\citeauthoryear{Weinberg and Blosser}{Weinberg and
  Blosser}{2009}]%
        {weinberg2009leader}
\bibfield{author}{\bibinfo{person}{Gil Weinberg} {and} \bibinfo{person}{Brian
  Blosser}.} \bibinfo{year}{2009}\natexlab{}.
\newblock \showarticletitle{A Leader-Follower Turn-Taking Model Incorporating
  Beat Detection in Musical Human-Robot Interaction}. In
  \bibinfo{booktitle}{\emph{Proceedings of the 4th {{ACM}}/{{IEEE}}
  International Conference on {{Human}} Robot Interaction}}
  \emph{(\bibinfo{series}{{{HRI}} '09})}. \bibinfo{publisher}{{Association for
  Computing Machinery}}, \bibinfo{address}{{New York, NY, USA}},
  \bibinfo{pages}{227--228}.
\newblock
\showISBNx{978-1-60558-404-1}
\urldef\tempurl%
\url{https://doi.org/10.1145/1514095.1514149}
\showDOI{\tempurl}


\bibitem[\protect\citeauthoryear{Welge and Hassenzahl}{Welge and
  Hassenzahl}{2016}]%
        {Welge2016}
\bibfield{author}{\bibinfo{person}{Julika Welge} {and} \bibinfo{person}{Marc
  Hassenzahl}.} \bibinfo{year}{2016}\natexlab{}.
\newblock \showarticletitle{Better than Human: {{About}} the Psychological
  Superpowers of Robots}.
\newblock \bibinfo{journal}{\emph{Lecture Notes in Computer Science (including
  subseries Lecture Notes in Artificial Intelligence and Lecture Notes in
  Bioinformatics)}}  \bibinfo{volume}{9979 LNAI} (\bibinfo{year}{2016}),
  \bibinfo{pages}{993--1002}.
\newblock
\showISBNx{9783319474366}
\showISSN{16113349}
\urldef\tempurl%
\url{https://doi.org/10.1007/978-3-319-47437-3_97}
\showDOI{\tempurl}


\bibitem[\protect\citeauthoryear{Wiese, Vallacher, and Strawinska}{Wiese
  et~al\mbox{.}}{2010}]%
        {nowak2007dynamical}
\bibfield{author}{\bibinfo{person}{Susan~L. Wiese}, \bibinfo{person}{Robin~R.
  Vallacher}, {and} \bibinfo{person}{Urszula Strawinska}.}
  \bibinfo{year}{2010}\natexlab{}.
\newblock \showarticletitle{Dynamical {{Social Psychology}}: {{Complexity}} and
  {{Coherence}} in {{Human Experience}}: {{Complexity}} and {{Coherence}} in
  {{Human Experience}}}.
\newblock \bibinfo{journal}{\emph{Social and Personality Psychology Compass}}
  \bibinfo{volume}{4}, \bibinfo{number}{11} (\bibinfo{date}{Nov.}
  \bibinfo{year}{2010}), \bibinfo{pages}{1018--1030}.
\newblock
\showISSN{17519004}
\urldef\tempurl%
\url{https://doi.org/10.1111/j.1751-9004.2010.00319.x}
\showDOI{\tempurl}


\bibitem[\protect\citeauthoryear{Yun, Watanabe, and Shimojo}{Yun
  et~al\mbox{.}}{2012}]%
        {yun2012interpersonal}
\bibfield{author}{\bibinfo{person}{Kyongsik Yun}, \bibinfo{person}{Katsumi
  Watanabe}, {and} \bibinfo{person}{Shinsuke Shimojo}.}
  \bibinfo{year}{2012}\natexlab{}.
\newblock \showarticletitle{Interpersonal Body and Neural Synchronization as a
  Marker of Implicit Social Interaction}.
\newblock \bibinfo{journal}{\emph{Scientific reports}} \bibinfo{volume}{2},
  \bibinfo{number}{1} (\bibinfo{year}{2012}), \bibinfo{pages}{1--8}.
\newblock
\urldef\tempurl%
\url{https://doi.org/10.1038/srep00959}
\showDOI{\tempurl}


\bibitem[\protect\citeauthoryear{Z{\"o}rner, Arts, Vasiljevic, Srivastava,
  Schmalzl, Mir, Bhatia, Strahl, Peters, Alpay, et~al\mbox{.}}{Z{\"o}rner
  et~al\mbox{.}}{2021}]%
        {zorner2021immersive}
\bibfield{author}{\bibinfo{person}{Sebastian Z{\"o}rner}, \bibinfo{person}{Emy
  Arts}, \bibinfo{person}{Brenda Vasiljevic}, \bibinfo{person}{Ankit
  Srivastava}, \bibinfo{person}{Florian Schmalzl}, \bibinfo{person}{Glareh
  Mir}, \bibinfo{person}{Kavish Bhatia}, \bibinfo{person}{Erik Strahl},
  \bibinfo{person}{Annika Peters}, \bibinfo{person}{Tayfun Alpay},
  {et~al\mbox{.}}} \bibinfo{year}{2021}\natexlab{}.
\newblock \showarticletitle{An Immersive Investment Game to Study Human-Robot
  Trust}.
\newblock \bibinfo{journal}{\emph{Frontiers in Robotics and AI}}
  \bibinfo{volume}{8} (\bibinfo{year}{2021}), \bibinfo{pages}{644529}.
\newblock
\urldef\tempurl%
\url{https://doi.org/10.3389/frobt.2021.644529}
\showDOI{\tempurl}


\end{thebibliography}
